\documentclass[AMA,STIX1COL]{WileyNJD-v2}

\usepackage{threeparttable}
\def \se {\text{se}}
\def \sp {\text{sp}}
\def \lnHR {\text{lnHR}}
\def \SROC {\text{SROC}}
\def \SAUC {\text{SAUC}}
\def \logit {\text{logit}}
\def \select {\text{select}}
\articletype{research article}%

\received{26 April 2016}
\revised{6 June 2016}
\accepted{6 June 2016}

\begin{document}

\title{Sensitivity analysis for publication bias on the time-dependent summary ROC analysis in meta-analysis of prognosis studies}

\author[1,3]{Yi Zhou}
\author[2]{Ao Huang}
\author[3,4]{Satoshi Hattori*}

\authormark{Zhou \textsc{et al.}}

\address[1]{\orgdiv{Beijing International Center for Mathematical Research}, \orgname{Peking University}, \orgaddress{\state{Beijing}, \country{China}}}
\address[2]{\orgdiv{Department of Medical Statistics}, \orgname{University Medical Center G{\"o}ttingen}, \orgaddress{\state{G{\"o}ttingen}, \country{Germany}}}
\address[3]{\orgdiv{Department of Biomedical Statistics, Graduate School of Medicine}, \orgname{Osaka University}, \orgaddress{\state{Osaka}, \country{Japan}}}
\address[4]{\orgdiv{Integrated Frontier Research for Medical Science Division, Institute for Open and Transdisciplinary Research Initiatives (OTRI)}, \orgname{Osaka University}, 
\orgaddress{\state{Osaka}, \country{Japan}}}

\corres{Satoshi Hattori, Department of Biomedical Statistics, Graduate School of Medicine, Osaka University, Osaka, Japan.\\
\email{hattoris@biostat.med.osaka-u.ac.jp}}

\abstract[Summary]{
In the analysis of prognosis studies with time-to-event outcomes, dichotomization of patients is often made. As the evaluations of prognostic capacity, the survivals of groups with high/low expression of the biomarker are often estimated by the Kaplan-Meier method, and the difference between groups is summarized via the hazard ratio (HR). The high/low expressions are usually determined by study-specific cutoff values, which brings heterogeneity over multiple prognosis studies and difficulty to synthesizing the results in a simple way. In meta-analysis of diagnostic studies with binary outcomes, the summary receiver operating characteristics (SROC) analysis provides a useful cutoff-free summary over studies. Recently, this methodology has been extended to the time-dependent SROC analysis for time-to-event outcomes in meta-analysis of prognosis studies. In this paper, we propose a sensitivity analysis method for evaluating the impact of publication bias on the time-dependent SROC analysis. Our proposal extends the recently introduced sensitivity analysis method for meta-analysis of diagnostic studies based on the bivariate normal model on sensitivity and specificity pairs. To model the selective publication process specific to prognosis studies, we introduce a trivariate model on the time-dependent sensitivity and specificity and the log-transformed HR. Based on the proved asymptotic property of the trivariate model, we introduce a likelihood based sensitivity analysis method based on the conditional likelihood constrained by the expected proportion of published studies. We illustrate the proposed sensitivity analysis method through the meta-analysis of Ki67 for breast cancer. Simulation studies are conducted to evaluate the performance of the proposed method.
}


\jnlcitation{\cname{%
\author{Y. Zhou}, 
\author{A. Huang}, and 
\author{S. Hattori}} (\cyear{2023}), 
\ctitle{Sensitivity analysis for publication bias on the time-dependent summary ROC analysis in meta-analysis of prognosis studies}, 
\cjournal{Stat Med.}, \cvol{2017;00:1--6}.}

\maketitle

\section{Introduction}
\label{sec1}

Biomarkers have been playing critical roles in medical therapeutics and precision medicine, and many clinical studies aim to investigate the associations between biomarkers and subjects' outcomes.
Biomarkers of strong associations with disease diagnosis and patients' prognosis are often referred as diagnostic and prognostic biomarkers, respectively.
Suppose we are interested in evaluating the association between a continuous biomarker and a time-to-event outcome in prognosis studies.
(Although prognosis studies could have continuous or binary outcomes, we focus on prognosis studies with time-to-event outcomes.)
To clarify some statements, we let diagnostic capacity be assessed by the association between subjects' biomarker expressions and their binary disease outcomes and prognostic capacity be assessed by that and their time-to-event outcomes.
Correspondingly, studies that evaluate the diagnostic or prognostic capacity of the biomarker of interest are called diagnostic or prognosis studies, respectively.

In the analysis of diagnostic or prognosis study with continuous biomarkers, subjects are often classified into the high/low expression (or positive/negative) groups by the study-specific cutoff value.
When the true disease and non-disease outcomes of subjects at diagnosis are known, diagnostic capacity of the biomarker is often represented by the sensitivity and specificity pair or the diagnostic odds ratio (DOR) at the specified cutoff value.
The receiver operator characteristic (ROC) analysis, including the ROC curve and the area under the curve (AUC), is more attractive since it presents the diagnostic capacity independent of a specific cut-off value.
For a prognosis study, at certain time $t$, time-to-event outcomes can be dichotomized into subjects' having an event prior to $t$ or surviving. 
Thus, sensitivity/specificity is extended into the time-dependent function as the probability of high/low expression of the biomarker conditional on subject's having an event prior to $t$/surviving, respectively.\cite{Heagerty2000}
Accordingly, the time-dependent ROC analysis is also developed for presenting prognostic capacity.\cite{Heagerty2000,Kamarudin2017}

In meta-analysis of diagnostic studies, study-specific cutoff values induce correlation between the empirical sensitivity and specificity pairs among multiple primary studies.
Thus, it is recommended to bivariately model the empirical sensitivity and specificity pairs and depict their relationship by the summary ROC (SROC) curve.\cite{Rutter2001,Reitsma2005,HaitaoChu2010}
The SROC curve presents the monotonic relationship between the sensitivity and $1-\text{specificity}$ at all possible cutoff values,
and the area under the SROC curve, namely the summary AUC (SAUC), gives a summary of diagnostic capacity. 
Properties and extensions of the SROC curve have been much discussed in many statistical literature, and the SROC analysis (the SROC curve and the SAUC) is recommended as main results in meta-analysis.\cite{Macaskill2022}
However, methods for meta-analysis of prognosis studies have been understudied.
Recently, the time-dependent SROC curve, denoted by SROC$(t)$, was proposed by modelling the empirical time-dependent sensitivity and specificity pairs.
To make inference about SROC$(t)$, Combescure et al.\cite{Combescure2016} employed the non-linear mixed model to model the biomarker and time-to-event distributions.
Hattori and Zhou\cite{Hattori2016} proposed the bivariate normal model and the bivariate binomial model by extending the bivariate models\cite{Rutter2001,Reitsma2005} for meta-analysis of diagnostic studies. 
Among the models estimating SROC$(t)$, the bivariate normal model of Hattori and Zhou (hereinafter, the HZ model) requires less data and is simpler in the implementation.
As proposed by Hattori and Zhou, the empirical time-dependent sensitivity and specificity pairs and the corresponding variances can be estimated by the retrieved number of patients and the Kaplan-Meier (KM) estimates at several time points.
The HZ model then bivariately model the empirical time-dependent sensitivity and specificity pairs to estimate SROC$(t)$ and the area under SROC$(t)$, denoted by SAUC$(t)$, for quantifying prognostic capacity.
By testing the null hypothesis that $H_0\text{: SAUC}(t)=0.5$, one could examine whether the biomarker has significant prognostic capacity.

The main advantage of the SROC/SROC$(t)$ analysis is that diagnostic/prognostic capacity is summarized without depending on a specific cutoff value. 
Same as in meta-analysis of intervention studies, publication bias (PB) can be a serious threat to the validity of estimates in the SROC/SROC$(t)$ analysis.
PB is induced by the phenomenon that studies with significant results are more likely to be selectively published.\cite{Sutton2000}
In meta-analysis of intervention studies, methods for assessing and adjusting PB have been intensively investigated. 
Despite of the popular graphical methods (e.g., the funnel plot and the trim-and-fill method), sensitivity analysis methods with selection functions, including the Heckman-type selection functions,\cite{Copas2000,Copas2001} the $t$-statistic based selection function,\cite{Copas2013} and the worst-case analysis,\cite{Copas2004} could give more reliable evaluation about the potential impact of PB.
In meta-analysis of diagnostic studies, some selection function based methods were proposed for dealing with PB on the SROC analysis.  
Most of these methods modeled the selective publication process by the Heckman-type selection functions.\cite{Hattori2018,Piao2019,Li2021}
More recently, Zhou et al.\cite{Zhou2022} introduced the cutoff-dependent selection function, which is the probit model on the $t$-type statistic of the empirical sensitivity and specificity pairs.
Specifically, the $t$-type statistic is defined by that of the linear combination of the logit-transformed empirical sensitivity and specificity pairs, including the $t$-statistic of the log-transformed DOR (lnDOR) as a special case.
The cutoff-dependent selection function can model a variety of selective publication processes determined by the significance of sensitivity, specificity, or both.
Since the scientific arguments in each primary study are mainly based on the cutoff-dependent quantities (e.g., sensitivity, specificity, or the lnDOR), the cutoff-dependent selection functions in Zhou et al. would be more appealing to model the selective publication.

In contrast to the recent progress on the sensitivity analysis methods for PB on the SROC analysis, no methods are available to handle PB on the SROC$(t)$ analysis. 
In this paper, we develop a sensitivity analysis method to evaluate the potential impact of PB on the SROC$(t)$ analysis based on the HZ model. 
In clinical prognosis literatures, the KM estimates of the high/low expression groups are often reported graphically with the $P$-values of the log-rank test and the hazard ratio (HR) highlighting the findings about prognostic capacity of the biomarker.
Thus, the significance of the log-rank test, or the significance of the log-transformed HR (lnHR) between two groups, would be responsible for the publication of prognosis studies, and then selection function on the test statistic would be more appealing than the Heckman-type selection function.
This motivated us to extend the sensitivity analysis method that employs the $t$-type statistic based selection function in meta-analysis of diagnostic studies\cite{Zhou2022} for prognosis studies. 
However, it is hard to directly apply the sensitivity analysis method of Zhou et al. to SROC$(t)$. 
In the method of Zhou et al., the $t$-type statistic can be written as the function of sensitivity and specificity, which are reported as outcomes in the primary studies; then, they successfully extended the likelihood based sensitivity analysis method of Copas\cite{Copas2013} by considering the bivariate normal distribution of the logit-transformed sensitivity and specificity. 
In prognosis studies, the time-dependent sensitivity and specificity pairs are not observable outcomes, and it is difficult to re-express the log-rank statistic or the $t$-statistic of lnHR as a function of them. 
{To overcome this, we considered to trivariately model the logit-transformed time-dependent sensitivity and specificity and the lnHR and proved the asymptotic distribution.}
Based on the trivariate model, the conditional likelihood taking into account the selection function is derived to make inference about the impact of PB on SROC$(t)$ and SAUC$(t)$.

The rest of this article is organized as follows. 
In Section \ref{sec2.0}, we describe the motivating meta-analysis of Ki67.
In Section \ref{sec2}, we introduce the data and review the HZ model without taking into account selective publication.
In Section \ref{sec3}, we propose the trivariate model and the sensitivity analysis method in details.
In Section \ref{sec4}, we revisit the meta-analysis of Ki67 and evaluate the potential impact of PB by the proposed method.
In Section \ref{sec5}, simulation studies were conducted to evaluate the performance of the proposed method.
In Section \ref{sec6}, we conclude with a discussion.

\section{Motivating example}
\label{sec2.0}

De Azambuja et al.\cite{DeAzambuja2007} conducted meta-analysis of prognosis studies by synthesizing the reported lnHRs to evaluate the association between Ki67 and the survival of patients in early breast cancer.
For the overall survival outcome, 38 studies reported the lnHRs with the corresponding variances.
The lnHR was defined by that between the high versus low expression of Ki67 groups and were mostly estimated by the Cox model\cite{Cox1972} in the primary studies,
The high/low expressions were decided by the study-specific cutoff values, ranging from 3.5\% to 28.6\%.
The synthesized HR was estimated as 1.95 with 95\% confidence interval (CI) to be 1.70–2.24.
Since the observed lnHRs were estimated based on heterogeneous cutoff values, the results of this meta-analysis contained much uncertainty.

Hattori and Zhou re-analyzed this meta-analysis by the HZ model and estimated SROC$(t)$ and SAUC$(t)$ of interest.
Among the 38 studies, they extracted the KM estimates of the high/low expression groups at several time points from 23 reported KM curves and estimated SROC$(t)$ with SAUC$(t)$ at the third ($t=3$) and the fifth ($t=5$) follow-up years.
A concern about the validity of their analysis is that the total 38 studies might not cover all the studies, known as the issue of selective publication.
Since the HZ model uses the KM estimates, the influence of not reporting the KM estimates should also be addressed, and such missingness is often referred as selective reporting. 
To address the validity of SROC$(t)$ and SAUC$(t)$ estimation, both selective publication and reporting bias should be taken with caution.
We refer to the bias due to the mixture of selective publication and selective reporting simply as PB.

\begin{figure}[!hbt]
\centering\includegraphics[width=1\columnwidth]{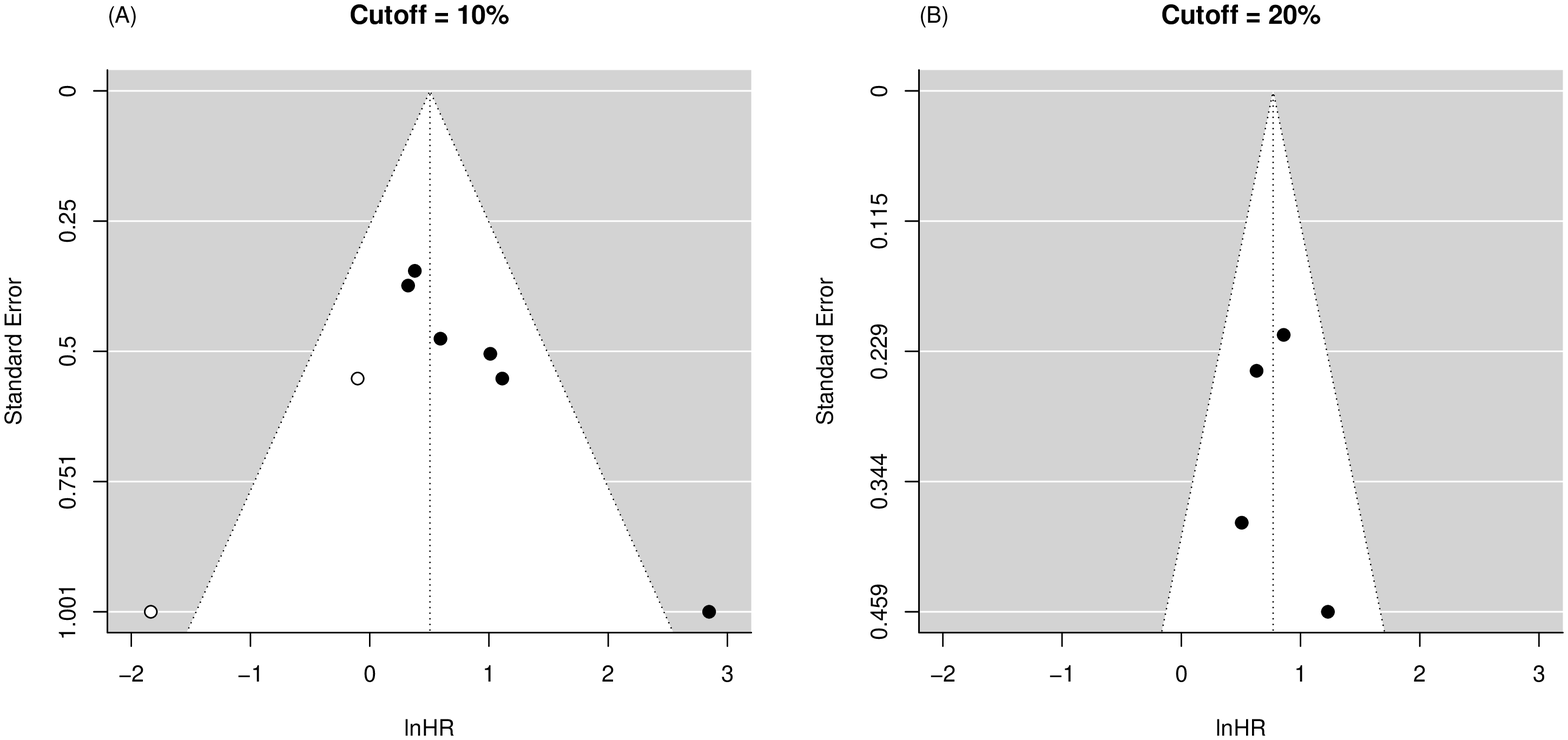}
\caption{The funnel plot and the trim and fill method for detecting PB in meta-analysis of Ki67. 
White points are the filled studies.}
\label{fig:ki67-tf}
\end{figure}

Since the lnHRs are dependent of the study-specific cutoff values, it would not be useful to see the funnel plot of the lnHR for all the studies. 
To exclude the influence of heterogeneous cutoff values, an intuitive idea is to apply the trim-and-fill method in the strata of cutoff values.
To roughly explore the potential of selective publication in this meta-analysis, we created the funnel plots on the subgroups of 23 lnHRs with common cutoff values, as shown in Figure \ref{fig:ki67-tf}.
In the subgroup of 10\% cutoff value (Figure \ref{fig:ki67-tf}A), the asymmetry of funnel plot suggested some potential of selective publication determined by the significance of lnHR; 
in the subgroup of 20\% cutoff value (Figure \ref{fig:ki67-tf}B), selective publication was not detected.
Although these funnel plots could not evaluate the potential impact of PB on the estimation of SROC$(t)$, they raised some concerns on PB, which motivated us to evaluate the impact of selective publication on SROC$(t)$ estimated by the HZ model.

\section{Bivariate normal model without considering publication bias}
\label{sec2}

\subsection{Notations and data}
\label{sec2.1}

Suppose that $N$ prognosis studies are published for the biomarker of interest, and there are $S$ studies totally including published and unpublished studies.
In this section, we do not consider any selective publication, that is, $N$ published studies are supposed to be all the existing studies or randomly selected from all the $S$ studies.
Considering to synthesize the published studies by the HZ model, we follow the notations in Hattori and Zhou to introduce the individual patient data (IPD) in each study and the meta-analytic data as follows.

Suppose each prognosis study $i\left(i = 1, 2, \dots, N\right)$ includes $n^{(i)}$ subjects, and the subjects are assumed to be random samples from the population of interest. 
For each subject, let $\tilde X$ be the baseline measurement of biomarker, $\tilde T$ the failure time, $\tilde C$ the right-censored time, and $\tilde Y = \mathrm{min}\left(\tilde T, \tilde C\right)$ the follow-up time.
Since the values of these variables are usually inaccessible for meta-analysis, we use tilde to denote the latent variables.
We assume that the distribution of $\tilde C$ is identical across the $S$ studies; in addition, $\tilde C {\perp\!\!\!\perp} \tilde X$, and $\tilde T {\perp\!\!\!\perp} \tilde C|\tilde X$. 
These assumptions lead to $\tilde T {\perp\!\!\!\perp} \tilde C$.
For study $i$, let $v^{(i)}$ denote the study-specific cutoff value, which is not necessarily reported.
The cutoff value separates subjects into the low expression group if $\tilde X \le v^{(i)}$ or the high expression group if $\tilde X > v^{(i)}$.
The survival functions of the low and high expression groups are defined by $S_0^{(i)}\left(t\right)=P\left(\tilde T > t | \tilde X \le v^{(i)}\right)$ and $S_1^{(i)}\left(t\right)=P\left(\tilde T > t| \tilde X > v^{(i)}\right)$, respectively.
The disease status at time $t$ is defined by the counting process $D(t)=1$ if $\tilde T\le t$ or $D(t)=0$ if $\tilde T > t$, indicating that subjects have an event before time $t$ or survive after $t$, respectively.
Sensitivity and specificity at time $t$ are defined by $\se\left(x,t\right) = P\left(\tilde X > x|\tilde T \le t\right)$ and $\sp\left(x,t\right) = P\left(\tilde X \le x|\tilde T > t\right)$, respectively.

The following data are observable for meta-analysis.
Let $n_0^{(i)}$ and $n_1^{(i)}$ denote the number of subjects separated into the low and high expression groups, respectively; the total number of subjects is $n^{(i)} = n_0^{(i)} + n_1^{(i)}$.
Let $\hat S_0^{(i)}\left(t\right)$ and $\hat S_1^{(i)}\left(t\right)$ denote the KM estimates of the low and high expression groups, respectively,
and can be extracted from the KM curves at the partition of time interval $\left[0,t\right]$: $0=t_0<t_1<\dots<t_K=t$.
Let $\hat \mu_{\lnHR}^{(i)}$ denote the reported lnHR between high versus low expression groups and $\hat s_{\lnHR}^{(i)}$ the standard error (SE); both are estimated by the Cox model.
In addition, some studies report the sample medians of the follow-up time over the total subjects.
The medians of follow-up time are used to estimate the censoring distribution used in the SEs of $\hat{S}_0^{(i)}(t)$ and $\hat{S}_1^{(i)}(t)$. (See Section 3.2 in Hattori and Zhou for more details.)

\subsection{Bivariate normal model}
\label{sec2.2}

After simple algebraic manipulations, the time-dependent sensitivity and specificity in study $i$ can be expressed by
\begin{equation}
\begin{aligned}
\text{Sensitivity:~}&\se\left(v^{(i)}, t \right) = \dfrac
{\left\{1- S_1^{(i)}\left(t\right)\right\}q_1^{(i)}}
{\left\{1- S_0^{(i)}\left(t\right)\right\}q_0^{(i)} + \left\{1- S_1^{(i)}\left(t\right)\right\}q_1^{(i)}}\\
\text{Specificity:~}&\sp\left(v^{(i)}, t\right) = \dfrac
{ S_0^{(i)}\left(t\right)\cdot q_0^{(i)}}
{ S_0^{(i)}\left(t\right)\cdot q_0^{(i)} + S_1^{(i)}\left(t\right)\cdot q_1^{(i)}},
\end{aligned}
\label{eq:sesp}
\end{equation}
where $q_0^{(i)} = P\left(\tilde X\le v^{(i)}\right)$ and $q_1^{(i)} = P\left(\tilde X>v^{(i)}\right)$ and are estimated by $\hat q_0^{(i)} = n_0^{(i)}/n^{(i)}$ and $\hat q_1^{(i)} = n_1^{(i)}/n^{(i)}$, respectively.
Sensitivity and specificity in equation (\ref{eq:sesp}) are consistently estimated by substituting $\left(\hat q_0^{(i)}, \hat q_1^{(i)}, \hat S_0^{(i)}, \hat S_1^{(i)}\right)$ for $\left(q_0^{(i)}, q_1^{(i)}, S_0^{(i)}, S_1^{(i)}\right)$,
and the resulting consistent estimators are denoted by $\hat\se\left(v^{(i)}, t \right)$ and $\hat\sp\left(v^{(i)}, t \right)$, respectively.

The HZ model is an extension of the bivariate random effects model of Reitsma et al.\cite{Reitsma2005} for survival outcomes.
Let $\boldsymbol \mu^{(i)} = \left(\mu_\se^{(i)},\mu_\sp^{(i)}\right)'$ denote the time-dependent sensitivity and specificity pair on the logit scale:
\begin{equation*}
\begin{aligned}
\mu_\se^{(i)}=\logit\left\{\se\left(v^{(i)},t\right)\right\}\text{~and~}
\mu_\sp^{(i)}=\logit\left\{\sp\left(v^{(i)},t\right)\right\}.
\end{aligned}
\label{eq:usesp}
\end{equation*}
At the between-study level, it is assumed that
\begin{align}
{\boldsymbol \mu}^{(i)}
\sim N_2 
\left (\boldsymbol \mu, \boldsymbol \Omega \right )
\text{~with~}
\boldsymbol \Omega=
\begin{bmatrix}
\tau_\se^2 & \tau_\mathrm{se,sp}\\ 
\tau_\mathrm{se,sp} & \tau_\sp^2
\end{bmatrix},
\label{eq:bnm2}
\end{align}
where $N_2$ denotes the bivariate normal distribution, $\boldsymbol \mu = \left(\mu_\se, \mu_\sp\right)'$ is the common mean at time $t$ across the prognosis studies,
and $\mathbf \Omega$ is the corresponding between-study variance-covariance matrix.

Let $\hat{\boldsymbol \mu}^{(i)} = \left(\logit\left\{\hat\se\left(v^{(i)},t\right)\right\}, \logit\left\{\hat\sp\left(v^{(i)},t\right)\right\}\right)'$ denote the consistent estimates of $\boldsymbol \mu^{(i)}$ in each study.
Hattori and Zhou showed that $\hat {\boldsymbol\mu}^{(i)}$ has the following asymptotic distribution at the within-study level:
\begin{align}
\hat{\boldsymbol \mu}^{(i)}| \boldsymbol \mu^{(i)}
\sim N_2 
\left (\boldsymbol \mu^{(i)}, \left.{ {\mathbf H}^{(i)}}\right/{n^{(i)}}  \right )
\text{~with~}
{\mathbf H}^{(i)} = 
\begin{bmatrix}
\left\{\sigma_\se^{(i)}\right\}^2 & \sigma_\mathrm{se,sp}^{(i)}\\ 
\sigma_\mathrm{se,sp}^{(i)} & \left\{\sigma_\sp^{(i)}\right\}^2
\end{bmatrix},
\label{eq:bnm1}
\end{align}
where ${\mathbf H}^{(i)}$ indicates the within-study asymptotic variance-covariance matrix.
See equation (\ref{eq:3sig}) in Appendix \ref{ap:A1} for more details of ${\mathbf H}^{(i)}$.
Following the convention that the within-study variance-covariance matrix is known in meta-analysis, we replace ${\mathbf H}^{(i)}$ in model (\ref{eq:bnm1}) with its consistent estimator $\hat{\mathbf H}^{(i)}$,  
which was estimated by the Greenwood formula and using median follow-up time. 
(See Section 3.2 in Hattori and Zhou for more details.)
Combining models (\ref{eq:bnm2}) and (\ref{eq:bnm1}) induces the HZ model:
\begin{align}
\hat{\boldsymbol \mu}^{(i)}
\sim N_2 
\left (\boldsymbol \mu, \left. \boldsymbol \Omega + {\hat{\mathbf H}^{(i)}}\right/{n^{(i)}} \right ),
\label{eq:bnm12}
\end{align}
where the parameters $(\boldsymbol \mu, \boldsymbol \Omega)$ are dependent of time $t$.

SROC$(t)$ is derived by taking the conditional expectation of $\mu_\se^{(i)}$ given $\mu_\sp^{(i)}$ in model (\ref{eq:bnm2}). 
Let $x$ denote $1-\sp\left(x,t\right)$, SROC$(t)$ is defined by the following time-dependent function:
\begin{align}
\SROC(t) = \SROC\left(x, t; \boldsymbol \mu, \mathbf \Omega\right) = \logit^{-1}\left[ \mu_\se - \dfrac{\tau_\mathrm{se,sp}}{\tau_\sp^2}\left\{ \logit(x)+\mu_\sp\right\}\right].
\label{eq:sroc}
\end{align}
Accordingly, SAUC$(t)$ is defined by
\begin{align}
\SAUC(t) = \SAUC\left(t; \boldsymbol \mu, \mathbf \Omega\right) = \int_0^1 \SROC\left(x, t; \boldsymbol \mu, \mathbf \Omega\right) dx.
\label{eq:sauc}
\end{align}


\section{Sensitivity analysis for publication bias} 
\label{sec3}

\subsection{Trivariate model incorporating the lnHR}
\label{sec3.0}

In practice, the publication of prognosis studies can be influenced by the $P$-value of the log-rank test.
The selective publication causes $N$ published studies to be biased sample from the $S$ studies and may induce PB in the estimates of $\SROC(t)$ and $\SAUC(t)$.
To evaluate the impact of PB on the SROC$(t)$ analysis, we need to expand the HZ model (\ref{eq:bnm12}) to correlate the time-dependent sensitivity and specificity with the lnHR.

To distinguish from notations for the HZ model (\ref{eq:bnm12}), 
we let $\hat {\boldsymbol y}^{(i)}=\left(\hat \mu_\se^{(i)},\hat \mu_\sp^{(i)}, \hat \mu_\lnHR^{(i)}\right)'$ denote the consistent estimators of 
$\boldsymbol \theta^{(i)} = \left( \mu_\se^{(i)},\mu_\sp^{(i)}, \mu_\lnHR^{(i)}\right)'$ at time $t$,
where $\mu_\lnHR^{(i)}$ denotes the lnHR and $\hat \mu_\lnHR^{(i)}$ the empirical lnHR estimated by the Cox model from the primary studies.
At the between-study level, it is assumed that 
\begin{align}
{\boldsymbol\theta}^{(i)} \sim N_3\left ({\boldsymbol\theta}, \mathbf \Psi \right )
~\text{with}~
\mathbf \Psi = 
\begin{bmatrix}
    \psi_\se^2    & \psi_\mathrm{se,sp}  & \psi_\mathrm{se,lnHR} \\ 
    \psi_\mathrm{se,sp}  & \psi_\sp^2    & \psi_\mathrm{sp,lnHR} \\ 
    \psi_\mathrm{se,lnHR} & \psi_\mathrm{sp,lnHR} & \psi_\lnHR^2
  \end{bmatrix},
\label{eq:tnm2}
\end{align}
where $\boldsymbol \theta = \left(\theta_\se, \theta_\sp, \theta_\lnHR\right)'$ are the common means, and $\mathbf \Psi$ indicates the between-study variance-covariance matrix.

At the within-study level, 
we proved that
\begin{align}
\hat{\boldsymbol y}^{(i)}| \boldsymbol\theta^{(i)} \sim N_3 \left ( \left.{\boldsymbol\theta^{(i)}}, {{\mathbf \Sigma}^{(i)}}\right/n^{(i)} \right )
\text{~with~}
{\mathbf \Sigma}^{(i)} = 
\begin{bmatrix}
\mathbf H^{(i)} & \mathbf \Sigma_{12}^{(i)}\\ 
{\mathbf \Sigma_{12}^{(i)}}' & \left\{ \sigma_\lnHR^{(i)}\right\}^2
\end{bmatrix},
\label{eq:tnm1}
\end{align}
where $N_3$ denotes the trivariate normal distribution, 
$\mathbf \Sigma^{(i)}$ the within-study asymptotic variance-covariance matrix,
${\mathbf H}^{(i)}$ the variance-covariance matrix in model (\ref{eq:bnm1});
{let $\mathbf \Sigma_{12}^{(i)}=\left(\sigma_\mathrm{se,lnHR}^{(i)}, \sigma_\mathrm{sp,lnHR}^{(i)}\right)'$, where  
$\sigma_\mathrm{se,lnHR}^{(i)}$ is the covariance between $\hat \mu_\se^{(i)}$ and $\hat \mu_\lnHR^{(i)}$ 
and $\sigma_\mathrm{sp,lnHR}^{(i)}$ the covariance between $\hat \mu_\sp^{(i)}$ and $\hat \mu_\lnHR^{(i)}$,
and $\left\{ \sigma_\lnHR^{(i)}\right\}^2$ the variance of $\hat \mu_\lnHR^{(i)}$.
The proof of the asymptotic distribution of $\hat{\boldsymbol y}^{(i)}$ (\ref{eq:tnm1}) is presented in Appendix \ref{ap:sigma}.
According to the convention in meta-analysis that the within-study variance-covariance matrix is known, $\mathbf \Sigma^{(i)}$ is replaced by its consistent estimator $\hat {\mathbf \Sigma}^{(i)}$.}
Combining models (\ref{eq:tnm1}) and (\ref{eq:tnm2}) gives the marginal distribution of $\hat {\boldsymbol y}^{(i)}$:
\begin{align}
\hat{\boldsymbol y}^{(i)}
\sim N_3 
\left (\boldsymbol \theta, \left. \boldsymbol \Psi + {\hat{\mathbf \Sigma}^{(i)}}\right/{n^{(i)}}  \right ).
\label{eq:tnm12}
\end{align}

\subsection{Selection functions on the significance of the lnHR}
\label{sec3.1}
From this section, we consider that the published $N$ studies are subject to selective publication in that a study with significant lnHR is more likely to be published.
The selective publication process is modeled by the probability of a study being selected given the $t$-statistic, as shown in the following selection function:
\begin{align}
P\left(\select\left| \hat{\boldsymbol y}^{(i)}, \hat {\mathbf \Sigma}^{(i)}\right.\right)
= a\left({\hat{\boldsymbol y}^{(i)}, \hat {\mathbf \Sigma}^{(i)}} \right)=a\left(t^{(i)}\right),
\label{eq:sfa}
\end{align}
where the function $a(\cdot)$ is the selection function. 
To simplify the inference procedure, we employ the probit model for $a(\cdot)$.
Thus, equation (\ref{eq:sfa}) is defined by:
\begin{align}
a\left(t^{(i)}\right)=\Phi\left(\alpha + \beta\cdot t^{(i)}\right),
\label{eq:sfa2}
\end{align}
where $t^{(i)} = \hat\mu_\lnHR\left/\hat s^{(i)}_\lnHR\right.$, parameters $\alpha$ and $\beta$ control the probability of selective publication,
and $\hat s^{(i)}_{\lnHR}$ denotes the reported SE with $\hat s^{(i)}_{\lnHR} = \left.\hat \sigma_\lnHR^{(i)}\right/\sqrt{n^{(i)}}$.
The monotonic property of the probit model links two cases of randomly selective publication:
(1) when $\beta=0$ and suppose $\alpha =\Phi^{-1}\left(p_0\right)$, 
the probability of selective publication is independent of the $t$-statistic, and each study is randomly published from the population with $P(\select)=p_0$;
(2) when $\beta\rightarrow \infty$, each study is published with $P(\select)=1$.

According to the definition of the probit model, the selection function (\ref{eq:sfa2}) can be represented by
\begin{align}
a\left(t^{(i)}\right)
=\Phi\left(z^{(i)} < \alpha + \beta\cdot t^{(i)}\right)= \Phi\left\{ \alpha + \beta\cdot \left(\left.{\hat \mu^{(i)}_\lnHR}\right/{\hat s^{(i)}_\lnHR}\right) \right\},
\label{eq:sfa3}
\end{align}
where $z^{(i)}$ is the standard normal random variable independent of $t^{(i)}$.
According to the trivariate model (\ref{eq:tnm12}), the marginal distribution of $\hat \mu^{(i)}_\lnHR$ is
\begin{align*}
\hat \mu^{(i)}_\lnHR\sim N\left(\theta^{(i)}_\lnHR, \psi_\lnHR^2 + \left\{\hat s^{(i)}_{\lnHR}\right\}^2 \right).
\end{align*}
Thus, the distribution of $t^{(i)}$ is derived by
\begin{align*}
N \left( \dfrac{\theta_\lnHR}{\hat s^{(i)}_\lnHR}, 1 + \left\{\dfrac{\psi_\lnHR}{\hat s^{(i)}_{\lnHR}}\right\}^2\right ),
\end{align*}
and the selection function $a\left(t^{(i)}\right)$ in equation (\ref{eq:sfa3}) can be written into the selection function $b\left(\hat{\boldsymbol\Sigma}^{(i)}\right)$:
\begin{align}
P\left(\select\left|{\hat {\mathbf \Sigma}^{(i)}}\right.\right) =
b\left(\hat{\boldsymbol \Sigma}^{(i)}\right) 
= \Phi\left\{
\dfrac{\alpha + \beta\cdot \left(\theta_\lnHR\left/\hat s^{(i)}_\lnHR\right.\right)  }
{\sqrt{1 + \beta^2\cdot \left\{1+ \left({\psi_\lnHR}\left/{\hat s^{(i)}_{\lnHR}}\right.\right)^2\right\}}}
\right\}.
\label{eq:sfb}
\end{align}

\subsection{Likelihood based sensitivity analysis}
\label{sec3.2}

Following the idea by Copas\cite{Copas2013} and Zhou et al.,\cite{Zhou2022}
we estimate the parameters in SROC$(t)$ and SAUC$(t)$ by maximizing the loglikelihood subject to a given marginal selection function, $P(\select)$, where $P(\select) = E_P\left\{a\left(t^{(i)}\right)\right\} = E_P\left\{b\left(\hat{\boldsymbol \Sigma}^{(i)}\right)\right\}$. 
We regard $P(\select)=p$ as the sensitivity parameter; with various values of $p$, the change of estimates is examined.
In this section, we derive the loglikelihood function of the published studies given a fixed value of $p$.

We let $f_P\left(\hat{\boldsymbol y}^{(i)}\left| \hat {\mathbf \Sigma}^{(i)}\right.\right)$ denote the marginal distribution of $\hat{\boldsymbol y}^{(i)}$ and $f_P\left(\hat {\mathbf \Sigma}^{(i)}\right)$ the distribution of $\hat {\mathbf \Sigma}^{(i)}$.
In meta-analysis without taking into account PB, the empirical data $\left(\hat{\boldsymbol y}^{(i)}, \hat {\mathbf \Sigma}^{(i)}\right)$ are regarded as random sample from the population with the joint distribution $f_P\left(\hat{\boldsymbol y}^{(i)}, \hat {\mathbf \Sigma}^{(i)}\right)$.
With selective publication, the published studies can be biased sample from the population; we let $f_O$ denote the distribution of the selectively published samples.
Given the fixed $p$, the distribution of $\hat{\mathbf\Sigma}^{(i)}$ in the published studies can be derived by
\begin{align*}
f_O\left(\hat{\mathbf\Sigma}^{(i)}\right) = f\left(\left.\hat{\mathbf\Sigma}^{(i)}\right|\select\right) 
= \dfrac{P\left(\select\left|\hat{\mathbf\Sigma}^{(i)}\right.\right)f_P\left(\hat{\mathbf\Sigma}^{(i)}\right)}{P\left(\select\right)} 
= \dfrac{b\left(\hat{\boldsymbol \Sigma}^{(i)}\right)f_P\left(\hat{\mathbf\Sigma}^{(i)}\right)}{p},
\end{align*}
which gives 
\begin{align}
f_P\left(\hat{\mathbf\Sigma}^{(i)}\right)= p\cdot \left\{b\left( \hat{\boldsymbol \Sigma}^{(i)}\right)\right\}^{-1}f_O\left(\hat{\mathbf\Sigma}^{(i)}\right).
\label{eq:fos2}
\end{align}
The joint distribution of the empirical data is derived by
\begin{align*}
f_O\left(\hat{\boldsymbol y}^{(i)}, \hat{\mathbf{\Sigma}}^{(i)}\right) 
& = f\left(\left.\hat{\boldsymbol y}^{(i)}, \hat{\mathbf{\Sigma}}^{(i)}\right|\select\right) \\
& = \dfrac{P\left(\select\left|\hat{\boldsymbol y}^{(i)}\right., \hat{\mathbf{\Sigma}}^{(i)} \right) f_P\left(\hat{\boldsymbol y}^{(i)}, \hat{\mathbf{\Sigma}}^{(i)}\right)}{p} \\
& = \dfrac
{a\left(t^{(i)}\right) f_P\left(\hat{\boldsymbol y}^{(i)} \left| \hat{\mathbf{\Sigma}}^{(i)}\right.\right) f_O\left(\hat{\mathbf{\Sigma}}^{(i)}\right)}
{b \left(\hat{\boldsymbol \Sigma}^{(i)}\right) }.
\end{align*}
This joint distribution allows us to derive the loglikelihood of published studies:
\begin{equation}
\begin{aligned}
\ell_O\left(\boldsymbol\theta, \boldsymbol{\Psi}, \alpha, \beta\right) 
&= \log\prod_{i=1}^N f_O\left(\hat{\boldsymbol{\mu}}^{(i)}, \hat{\mathbf{\Sigma}}^{(i)}\right) \\
&= \sum_{i=1}^{N}\log f_P\left(\hat{\boldsymbol{\mu}}^{(i)}\left| \hat{\mathbf{\Sigma}}^{(i)}\right.\right)
+  \sum_{i=1}^{N}\log a\left(t^{(i)}\right) 
- \sum_{i=1}^{N}\log b\left(\hat{\mathbf{\Sigma}}^{(i)}\right)
+ c,
\end{aligned}
\label{eq:llk1}
\end{equation}
where $c = \sum_{i=1}^{N}\log\left\{ {f_O\left(\hat{\mathbf{\Sigma}}^{(i)}\right)}\right\}$ is constant.

Noting that by taking the integral of the both sides in equation (\ref{eq:fos2}), we can derive 
\begin{align}
p = 1\left/E_O\left\{b\left(\hat{\mathbf{\Sigma}}^{(i)}\right)^{-1}\right\}\right. \simeq 1\left/\sum_{i=1}^N b\left(\hat{\mathbf{\Sigma}}^{(i)}\right)^{-1}\right.,
\label{eq:p}
\end{align}
{where the function $b(\cdot)$ defined in equation (\ref{eq:sfb}) allows the parameter $\alpha$ to be represented by the function of $(\boldsymbol\theta, \boldsymbol{\Psi}, \beta)$ given a value of $p$; }
we denote this by $\alpha_p = \alpha_p(\boldsymbol\theta, \boldsymbol{\Psi}, \beta)$.
By replacing the $\alpha$ with $\alpha_p$, 
the loglikelihood (\ref{eq:llk1}) finally derives the conditional loglikelihood:
\begin{equation}
\begin{aligned}
\ell_O\left(\boldsymbol\theta, \boldsymbol{\Psi}, \beta; p\right) 
&\propto -\dfrac{1}{2}\sum_{i=1}^{N}\left\{  
\left(\hat{\boldsymbol y}^{(i)} - \boldsymbol\theta\right)' 
\left(\boldsymbol{\Psi}+\left.\hat{\mathbf{\Sigma}}^{(i)}\right/n^{(i)}\right)^{-1}
\left(\hat{\boldsymbol y}^{(i)}- \boldsymbol{\theta}\right) 
+ \log\left|\boldsymbol{\Psi}+\left.\hat{\mathbf{\Sigma}}^{(i)}\right/n^{(i)}\right|\right \} \\
& +  \sum_{i=1}^{N}\log \Phi\left\{ \alpha_p + \beta\cdot \left({\hat \mu^{(i)}_\lnHR}\left/{\hat s^{(i)}_\lnHR}\right.\right) \right\} 
- \sum_{i=1}^{N}\log \Phi\left\{
\dfrac{\alpha_p + \beta\cdot \left(\theta_\lnHR\left/\hat s^{(i)}_\lnHR\right.\right)  }
{\sqrt{1 + \beta^2\cdot \left\{1+ \left({\psi_\lnHR}\left/{\hat s^{(i)}_{\lnHR}}\right)^2\right.\right\}}}
\right\}.
\end{aligned}
\label{eq:llk2}
\end{equation}
The parameters in the loglikelihood (\ref{eq:llk2}) can be estimated by maximizing this conditional loglikelihood {given a prespecified $p$.}
We denote these maximum likelihood estimates (MLEs) to be $(\hat{\boldsymbol\theta}, \hat{\boldsymbol\Psi}, \hat\beta)$.
With the MLEs, SROC$(t)$ in equation (\ref{eq:sroc}) and SAUC$(t)$ in equation (\ref{eq:sauc}) can be estimated accordingly, 
denoted by $\mathrm{SR\hat OC}(t) =\SROC(x, t; \hat{\boldsymbol \theta}, \hat{\mathbf \Psi})$ and $\mathrm{SA\hat UC}(t) = \SAUC(t; \hat{\boldsymbol \theta}, \hat{\mathbf \Psi})$, respectively. 
The asymptotic normality of $(\hat{\boldsymbol\theta}, \hat{\boldsymbol\Psi}, \hat\beta)$ follows the general theory of the maximum likelihood estimation under the assumptions that $S$ and $n^{(i)}$ are large.
The asymptotic variance-covariance matrix of $(\hat{\boldsymbol\theta}, \hat{\boldsymbol\Psi}, \hat\beta)$ can be consistently estimated by the inverse of the empirical Fisher information following the maximum likelihood theory.
The two-tailed CI for the SAUC can be constructed by the delta method,
and the detailed derivation is presented in Appendix \ref{ap:sauc}. 
In practice, the true value of $p$ is unknown.
Thus, it is recommended to specify a range of values on $p$ to examine the changes of $\mathrm{SR\hat OC}(t)$ or $\mathrm{SA\hat UC}(t)$ as sensitivity analysis.

\section{Application}
\label{sec4}

\begin{figure}[!hbt]
\centering\includegraphics[width=1\columnwidth]{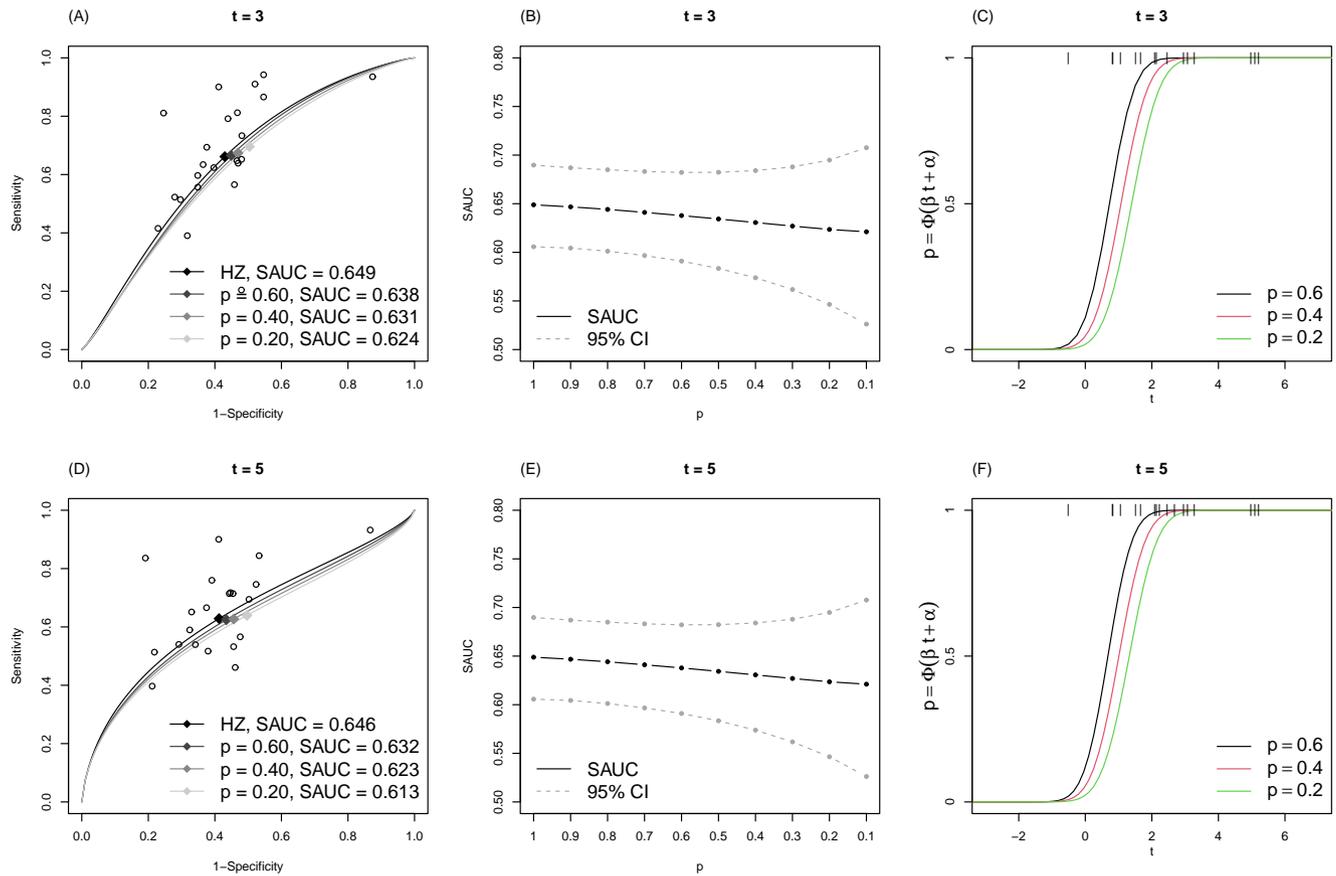}
\caption{
The estimated SROC$(t)$, SAUC$(t)$, and the probit selection function at $t=3, 5$ when $p=0.6, 0.4, 0.2$ in Ki-67 example.
In panel (A) and (D), the circle points are the empirical $\mathrm{se}(x,t)$ and $1-\mathrm{sp}(x,t)$ pairs from 23 prognostic studies;
the diamond points are the estimated summary operating points, $\left(\mathrm{logit}^{-1}\hat\mu_{\mathrm{se}}, 1-\mathrm{logit}^{-1}\hat\mu_{\mathrm{sp}}\right)$.
Panel (B) and (E) show the estimated SAUC$(t)$ by the HZ model ($p=1$) and the proposed method given $p=0.9, \dots, 0.1$.
In panel (C) and (F), the vertical lines at the top are the observed $t$-statistics from 23 prognostic studies.
}
\label{fig:ki67}
\end{figure}

We revisited the meta-analysis of Ki67\cite{DeAzambuja2007} to evaluate the potential impact of PB on SROC$(t)$ and SAUC$(t)$ at $t=3$ and $t=5$.
As aforementioned, 38 studies reported the lnHRs for the overall survival outcome.
When $t=3$ and $t=5$, a total of 23 and 21 studies, respectively, reported the evaluable KM estimates.
Thus, we presumed the $P(\select)$ to be $p<23/38\approx 0.6$ and implemented the proposed sensitivity analysis on SROC$(t)$ analysis given $p=0.6, 0.4, 0.2$.
We regarded the HZ model  without taking into account PB (i.e., $p=1$) as the benchmark.
The changes of SROC$(t)$ at $t=3$ and $t=5$ are shown in Figure \ref{fig:ki67}A and \ref{fig:ki67}D, respectively.
The HZ model estimated SAUC$(t)$ at  $t=3$ and $t=5$ to be 0.649 (95\% CI: 0.606, 0.690) and 0.646 (0.610, 0.680), respectively.
When $p=0.6$, indicating that $(1-0.6)/0.6\times23 \approx 15$ unpublished studies potentially exist, the estimated SAUC$(t)$ at  $t=3$ and $t=5$ decreased to 0.638 (0.591, 0.682) and 0.632 (0.592, 0.677), respectively.
In the worse case when $p=0.2$, the estimated SAUC$(t)$ at $t=3$ and $t=5$ decreased to 0.624 (0.546, 0.695) and 0.613 (0.545, 0.677), respectively.

The changes of the estimated SAUC$(t)$ at $t=3$ and $t=5$ given $p=1$ (i.e., the~HZ~model), $0.9, \dots, 0.1$ are shown in Figure \ref{fig:ki67}B and \ref{fig:ki67}E, respectively.
Although SAUC$(t)$ decreased with decreasing $p$, the estimates of SAUC$(t)$ were still significantly different from 0.5.
The estimated probit selection functions (equation \ref{eq:sfa2}) at $t=3$ and $t=5$ were presented in Figure \ref{fig:ki67}C and \ref{fig:ki67}F, respectively.
The $t$-statistics of the published studies were shown as the vertical lines, and most studies had high probability of being selected.
The MLEs of the parameters were presented in Table S1 in the Supplementary Material.
The detailed estimates of SAUC$(t)$ in Figure \ref{fig:ki67}B and \ref{fig:ki67}E were presented in Table S2 in the Supplementary Material.

Although SAUC$(t)$ was not high, the prognostic capacity of Ki67 was presented to be statistically significant in Hattori and Zhou.
The sensitivity analysis indicated SAUC$(t)$ estimates were affected by PB to a small degree at the third and the fifth years.
After the sensitivity analysis, one could draw robust conclusion that the prognostic capacity of Ki67 antigen was not very high in patients with early breast cancer.


\section{Simulation studies}
\label{sec5}
 
Simulation studies were conducted to evaluate the performance of the proposed sensitivity analysis method.
We generated prognosis studies with the IPD and used the observable data mentioned in Section \ref{sec2.1} for meta-analysis. 
In each prognosis study $i$, we considered the following scenarios to generate the IPD.\cite{Hattori2016}
We considered one moderate size of total subjects and one large size with $n^{(i)}$ generated from the uniform distribution $U(50,150)$ and $U(50, 300)$, respectively.
The failure time of each subject $\tilde T$ was generated from $\log \tilde T = 1+\epsilon$, where $\epsilon$ followed the standard normal distribution.
The potential censored time $\tilde C$ was generated from the exponential distribution with hazard rate $\lambda = 0.2$, denoted by $Exp(0.2)$.
The asymptotic variance-covariance matrices in equations (\ref{eq:bnm12}) and (\ref{eq:tnm12}) required the limiting variances of $\hat S_0(t)$ and $\hat S_1(t)$, and the distribution of $\tilde C$ needed to be estimated.
Assuming that $\tilde C$ had the exponential distribution, we followed the method of Hattori and Zhou and estimated $\lambda$ by using the reported medians of the follow-up time.
To evaluate the situation when the distribution of $\tilde C$ was misfitted, we considered to generate the true $\tilde C$ from $U(1,4)$ additionally and misspecified the exponential distribution with the estimated $\lambda$ to fit $\tilde C$. 
Suppose we were interested in estimating SAUC$(t)$ at $t=2$, denoted by SAUC(2).
Two scenarios of the biomarker $\tilde X$ were considered:
\begin{equation*}
\begin{aligned}
\text{Biomarker1: }\tilde X = 
\left\{\begin{matrix}
0.7+0.1e &\mathrm{if~} \tilde T\le2
\\ 
0.3+0.3e &\mathrm{if~} \tilde T>2
\end{matrix}\right.
\end{aligned} 
\end{equation*}
\begin{equation*}
\begin{aligned}
\text{Biomarker2: }\tilde X = 
\left\{\begin{matrix}
0.6+0.2e &\mathrm{if~} \tilde T\le2
\\ 
0.4+0.3e &\mathrm{if~} \tilde T>2
\end{matrix}\right.
\end{aligned} 
\end{equation*}
where $e$ followed the standard logistic distribution. 
Biomarker1 induced comparatively large lnHR with $\mu_{\lnHR} \approx 0.96$ and Biomarker2 induced smaller lnHR with $\mu_{\lnHR}\approx 0.36$ when $\tilde C$ was generated from $Exp (0.2)$;
Biomarker1 induced $\mu_{\lnHR} \approx 1.36$ and Biomarker2 induced $\mu_{\lnHR}\approx 0.48$ when $\tilde C$ was generated from $U(1,4)$.
The cutoff value $v^{(i)}$ was generated from the normal distribution $N(0.5, 0.1^2)$.
With the IPD in study $i$, the sensitivity and specificity at $t=2$ were derived according to equation (\ref{eq:sesp}), and $\hat\mu_{\lnHR}^{(i)}$ was estimated by the Cox model on the biomarker $\tilde X$.
The asymptotic variance-covariance matrix in the trivariate model (\ref{eq:tnm2}) was derived according to equations (\ref{eq:3sig}), (\ref{eq:sig13}), and (\ref{eq:sig23}) in Appendix \ref{ap:sigma}.

{For meta-analysis, we considered two population sizes with $S=70$ and $200$.}
$P(\select)$ was set to be $p = 0.7,0.5,0.3$, which induced the number of published studies to be $N=S\times p$, ranging from 21 to 140.
In each selective publication process, we generated $S$ population studies and then selected $N$ studies according to the probit selection function (equation \ref{eq:sfa2}). 
In the true probit selection function, $\beta$ was set to be 1, and $\alpha$ at $p=0.7, 0.5, 0.3$ was calculated by simulation. The corresponding values of $\alpha$ were presented in Table S3 in the Supplementary Material.
The publication process was repeated 1000 times in each scenario.

We estimated SAUC(2) by maximizing the likelihood of the HZ model based on $S$ population studies and $N$ published studies, denoted by HZ$_P$ and HZ$_O$, respectively. 
SAUC(2) was also estimated by the proposed method based on $N$ published studies, denoted by Prop$_{(p)}$ with $p\approx N/S$. 
The assignment of $p$ was practically infeasible but was used as the specified value for evaluating the performance of the proposed method.
The estimated SAUC(2) by the HZ$_P$ method was regarded as the reference, and the difference between the estimates of HZ$_P$ and HZ$_O$ indicated the impact of PB.
The performance of the proposed method was assessed by the differences from the estimated SAUC(2) by the HZ$_P$ method.
All statistical computing was conducted by R (R Development Core Team, Version 4.1.3).
The maximum likelihood estimations were conducted by R function \texttt{nlminb()} with initial values being the MLEs of the trivariate model based on the published studies.
The lnHR was estimated by the Cox model conducted by the R package \texttt{survival}.

The medians with the 25th and 75th empirical percentiles of the estimated SAUC(2) were summarized in Table \ref{tab:tab1}-\ref{tab:tab2}.
The convergence rates (CRs) of all the methods were also presented; the CR was calculated by the proportion of successfully obtaining the converged estimates among 1000 repetitions.
When the censoring distribution was correctly fitted, the summaries of the SAUC(2) from Biomarker1 and Biomarker2 were presented in Table \ref{tab:tab1}.
When the lnHR was large (Biomarker1), the referential SAUC(2) was estimated to be approximately 0.75 by the HZ$_P$ method.
The increasing PB could be observed with decreasing $p$, while the proposed method reduced the bias and derived the estimates of SAUC(2) close to those of the HZ$_P$ method. 
When the lnHR was small (Biomarker2), the SAUCs were estimated to be approximately 0.62 by the HZ$_P$ method, and the proposed methods also reduced PB.
Since the proposed method is based on the trivariate model, one limitation of the proposed method was the lower CR compared with the HZ model. The simulation results indicated that the convergence rates became lower when lnHR was smaller.

When the distribution of the censored time was incorrectly fitted by the exponential distribution, the summaries in Table \ref{tab:tab2} implied that the proposed method could reduce PB. The results were in agreement with those when the censoring distribution was correctly fitted; however, the misspecification of the censoring distribution reduced the CRs.


\section{Discussion}
\label{sec6}

While the meta-analysis of diagnosis studies is widely applied in practice and indeed the Cochrane collaboration provides a nice introductory and educational summary of the applicable methodologies.\cite{Macaskill2022} 
Among them, the SROC analysis would be one of the most appealing methods in the presence of heterogeneous cutoff values. 
On the other hand, less attention was paid to meta-analysis of prognosis studies.
{As prognostic capacity is mostly measured by the HR (or the lnHR) from the Cox model, some meta-analytic methods were proposed to synthesize the HRs.
Riley et al.\cite{Riley2015} applied a meta-regression model by using the cutoff values as the explanatory variable.
Sadashima et al.\cite{Sadashima2016} summarized HRs with study-specific cutoff values by estimating an underlying individual-level model.
Alternative to the HR based meta-analytic methods, the recently developed time-dependent SROC analysis presents the time-dependent summary of results from studies with survival outcomes in the presence of heterogeneous cutoff values.\cite{Combescure2016,Hattori2016} 
However, PB may distort the meta-analytic estimates.
In meta-analysis of diagnostic studies, several selection function based methods have been successfully proposed to deal with PB,\cite{Hattori2018,Piao2019,Li2021} including the sensitivity analysis methods.\cite{Zhou2022}}
In meta-analysis of prognosis studies with time-to-event outcomes, to our best knowledge, no formal sensitivity analysis method is available for evaluating the impact of PB. 
Since the log-rank test and the HR between the high/low expression groups are widely used on time-to-event outcomes, the $P$-value or the test statistic would be responsible for the selective publication process, and selection functions defined on the test statistic would be appealing. 
{Thus, we employed the selection function on the $t$-statistic of lnHR and introduced the likelihood based sensitivity analysis on the trivariate model.}

Since the proposed conditional loglikelihood given the published is complex, the converged estimates may not be always obtained in some situations, where, for example, the number of published studies is small or the underlying censoring distribution is misfitted.
With enough number of published studies, the simulation studies showed that the proposed method could obtain CR over 70\%.
In practice, we suggest to try different plausible initial values such as some values that are speculated to be close to the true values of some parameters.
Despite of potential difficulty in finding the right initial values in implementation, our method provides the first method to address the PB issue in meta-analysis of prognosis studies.
We believe that this development weakens concerns in applying the SROC analysis to meta-analysis of prognosis studies, and the SROC methodologies should contribute to getting sound evidence from prognosis studies. 


\section*{Acknowledgments}

This research was partly supported by Grant-in-Aid for Challenging Exploratory Research (16K12403) and for Scientific Research(16H06299, 18H03208) from the Ministry of Education, Science, Sports and Technology of Japan.

\section*{Data available statement}
R codes together with a sample application data set are available at
\url{https://github.com/meta2020/progmetasa-r}.

\section*{Supporting information}
Additional supporting information can be found online in the Supporting Information section at the end of this article.

\bibliography{refs}%
\clearpage

\begin{table}[!htb]

\caption{\label{tab:tab1}Summary of the estimated SAUC when the true censoring is distributed as $Exp(0.2)$.}
\centering
\begin{threeparttable}
\begin{tabular}[t]{rrrrrrrr}
\toprule
\multicolumn{1}{c}{} & \multicolumn{1}{c}{} & \multicolumn{1}{c}{} & \multicolumn{1}{c}{} & \multicolumn{2}{c}{Biomarker1} & \multicolumn{2}{c}{Biomarker2} \\
\cmidrule(l{3pt}r{3pt}){5-6} \cmidrule(l{3pt}r{3pt}){7-8}
Patients & $S$ & $p$ & Methods & Median (Q1, Q3) & CR & Median (Q1, Q3) & CR\\
\midrule
50-150 & 70 &  & HZ$_P$ & 74.66 (73.70, 75.65) & 100 & 62.18 (61.55, 62.84) & 100\\
 &  & 0.7 & HZ$_O$ & 75.74 (74.59, 76.84) & 100 & 63.97 (63.33, 64.68) & 100\\
 &  &  & Prop$_{(0.7)}$ & 75.30 (73.90, 76.84) & 77.3 & 62.21 (61.34, 63.09) & 64\\
 &  & 0.5 & HZ$_O$ & 76.29 (74.80, 77.76) & 100 & 64.99 (64.07, 65.82) & 100\\
 &  &  & Prop$_{(0.5)}$ & 75.35 (73.51, 77.38) & 76.3 & 62.20 (60.92, 64.01) & 72\\
 &  & 0.3 & HZ$_O$ & 77.29 (75.29, 79.32) & 100 & 66.36 (65.14, 67.40) & 99.9\\
 &  &  & Prop$_{(0.3)}$ & 75.54 (72.86, 78.31) & 67.5 & 64.70 (60.89, 66.78) & 79.1\\
\addlinespace
 & 200 &  & HZ$_P$ & 74.51 (73.92, 75.13) & 100 & 62.22 (61.86, 62.58) & 100\\
 &  & 0.7 & HZ$_O$ & 75.54 (74.83, 76.19) & 100 & 64.03 (63.59, 64.43) & 100\\
 &  &  & Prop$_{(0.7)}$ & 75.00 (73.98, 75.94) & 72.2 & 62.08 (61.54, 62.56) & 63\\
 &  & 0.5 & HZ$_O$ & 76.15 (75.25, 76.93) & 100 & 65.06 (64.59, 65.57) & 100\\
 &  &  & Prop$_{(0.5)}$ & 75.01 (73.68, 76.26) & 76.3 & 61.75 (61.03, 62.50) & 67.2\\
 &  & 0.3 & HZ$_O$ & 76.80 (75.79, 77.85) & 100 & 66.30 (65.67, 66.96) & 100\\
 &  &  & Prop$_{(0.3)}$ & 74.81 (72.95, 76.73) & 76.8 & 62.04 (60.31, 66.21) & 73.1\\
\addlinespace
50-300 & 70 &  & HZ$_P$ & 75.75 (75.08, 76.48) & 100 & 62.53 (62.05, 62.96) & 100\\
 &  & 0.7 & HZ$_O$ & 76.52 (75.56, 77.47) & 100 & 63.81 (63.35, 64.37) & 100\\
 &  &  & Prop$_{(0.7)}$ & 76.23 (75.02, 77.27) & 77.4 & 63.15 (62.39, 63.86) & 73.9\\
 &  & 0.5 & HZ$_O$ & 76.98 (75.69, 78.17) & 100 & 64.59 (63.95, 65.26) & 100\\
 &  &  & Prop$_{(0.5)}$ & 76.18 (74.59, 77.54) & 74.8 & 63.45 (62.17, 64.50) & 66.2\\
 &  & 0.3 & HZ$_O$ & 77.32 (75.07, 79.13) & 100 & 65.51 (64.68, 66.31) & 99.9\\
 &  &  & Prop$_{(0.3)}$ & 75.67 (72.84, 78.46) & 75.7 & 64.43 (62.10, 65.79) & 70.1\\
\addlinespace
 & 200 &  & HZ$_P$ & 75.50 (75.07, 76.00) & 100 & 62.55 (62.25, 62.82) & 100\\
 &  & 0.7 & HZ$_O$ & 76.18 (75.62, 76.73) & 100 & 63.86 (63.55, 64.18) & 100\\
 &  &  & Prop$_{(0.7)}$ & 75.89 (75.23, 76.56) & 78.9 & 63.12 (62.51, 63.77) & 70.2\\
 &  & 0.5 & HZ$_O$ & 76.63 (75.87, 77.24) & 100 & 64.60 (64.25, 64.98) & 100\\
 &  &  & Prop$_{(0.5)}$ & 75.83 (74.99, 76.63) & 75.3 & 62.52 (61.97, 63.40) & 59.9\\
 &  & 0.3 & HZ$_O$ & 77.04 (76.01, 78.06) & 100 & 65.47 (65.04, 66.03) & 100\\
 &  &  & Prop$_{(0.3)}$ & 75.47 (74.19, 76.81) & 79.6 & 62.64 (61.58, 65.08) & 69.3\\
\bottomrule
\end{tabular}
\begin{tablenotes}
\item 
  Patients denote the range of the number of patients.
  $S$ denotes the number of the population studies.
  $p$ denotes the approximate proportion of the published studies among the population.
	Median with 25th and 75th empirical quartiles (Q1, Q3) of the SAUC at $t=2$ are reported. 
	CR denotes the proportion of successfully convergenced estimates among 1000 repetition.
	HZ$_P$ denotes the HZ model using the population studies; 
	HZ$_O$ denotes the HZ model using only the corresponding numbers of published studies;
	Prop$_{(p)}$ denotes the proposed sensitivity analysis method given $p$.
	All the entries are multiplied by 100.
\end{tablenotes}
\end{threeparttable}
\end{table}

\begin{table}[!htb]

\caption{\label{tab:tab2}Summary of the estimated SAUC when the true censoring is distributed as $U(1,4)$, but a misspecified exponential distribution is fitted.}
\centering
\begin{threeparttable}
\begin{tabular}[t]{rrrrrrrr}
\toprule
\multicolumn{1}{c}{} & \multicolumn{1}{c}{} & \multicolumn{1}{c}{} & \multicolumn{1}{c}{} & \multicolumn{2}{c}{Biomarker1} & \multicolumn{2}{c}{Biomarker2} \\
\cmidrule(l{3pt}r{3pt}){5-6} \cmidrule(l{3pt}r{3pt}){7-8}
Patients & $S$ & $p$ & Methods & Median (Q1, Q3) & CR & Median (Q1, Q3) & CR\\
\midrule
50-150 & 70 &  & HZ$_P$ & 74.99 (74.12, 75.80) & 100 & 62.28 (61.71, 62.82) & 100\\
 &  & 0.7 & HZ$_O$ & 75.90 (74.88, 77.02) & 100 & 64.25 (63.61, 64.92) & 100\\
 &  &  & Prop$_{(0.7)}$ & 75.57 (74.23, 77.04) & 63.9 & 62.56 (61.79, 63.41) & 44\\
 &  & 0.5 & HZ$_O$ & 76.57 (75.20, 77.97) & 100 & 65.35 (64.61, 66.19) & 100\\
 &  &  & Prop$_{(0.5)}$ & 75.02 (73.40, 76.78) & 56.8 & 62.94 (61.31, 65.20) & 51.5\\
 &  & 0.3 & HZ$_O$ & 77.31 (75.46, 79.32) & 100 & 66.70 (65.71, 67.75) & 100\\
 &  &  & Prop$_{(0.3)}$ & 74.65 (71.97, 77.27) & 45.4 & 66.26 (62.86, 67.51) & 64.4\\
\addlinespace
 & 200 &  & HZ$_P$ & 74.75 (74.17, 75.26) & 100 & 62.26 (61.92, 62.61) & 100\\
 &  & 0.7 & HZ$_O$ & 75.74 (75.13, 76.39) & 100 & 64.23 (63.88, 64.62) & 100\\
 &  &  & Prop$_{(0.7)}$ & 75.41 (74.50, 76.38) & 79.6 & 62.23 (61.72, 62.73) & 39.2\\
 &  & 0.5 & HZ$_O$ & 76.27 (75.48, 77.05) & 100 & 65.41 (64.98, 65.82) & 100\\
 &  &  & Prop$_{(0.5)}$ & 74.72 (73.57, 75.94) & 72.9 & 61.94 (61.30, 62.84) & 45.8\\
 &  & 0.3 & HZ$_O$ & 76.92 (75.85, 77.98) & 100 & 66.71 (66.07, 67.28) & 100\\
 &  &  & Prop$_{(0.3)}$ & 73.96 (72.54, 75.53) & 67.9 & 65.94 (61.18, 66.96) & 59.6\\
\addlinespace
50-300 & 70 &  & HZ$_P$ & 76.03 (75.28, 76.70) & 100 & 62.55 (62.11, 62.93) & 100\\
 &  & 0.7 & HZ$_O$ & 76.77 (75.78, 77.68) & 100 & 63.95 (63.50, 64.43) & 100\\
 &  &  & Prop$_{(0.7)}$ & 76.26 (75.13, 77.37) & 62.3 & 62.73 (62.08, 63.38) & 40.6\\
 &  & 0.5 & HZ$_O$ & 77.17 (75.76, 78.42) & 100 & 64.83 (64.23, 65.37) & 100\\
 &  &  & Prop$_{(0.5)}$ & 75.81 (74.31, 76.98) & 58.1 & 63.10 (61.88, 64.49) & 41.5\\
 &  & 0.3 & HZ$_O$ & 77.41 (74.34, 79.47) & 100 & 65.79 (64.98, 66.45) & 100\\
 &  &  & Prop$_{(0.3)}$ & 73.57 (34.02, 76.50) & 54.7 & 64.95 (62.12, 66.19) & 50.6\\
\addlinespace
 & 200 &  & HZ$_P$ & 75.78 (75.33, 76.18) & 100 & 62.59 (62.33, 62.86) & 100\\
 &  & 0.7 & HZ$_O$ & 76.40 (75.88, 76.93) & 100 & 63.97 (63.69, 64.29) & 100\\
 &  &  & Prop$_{(0.7)}$ & 76.18 (75.56, 76.77) & 72.5 & 62.60 (62.24, 62.95) & 34.6\\
 &  & 0.5 & HZ$_O$ & 76.83 (76.10, 77.52) & 100 & 64.80 (64.46, 65.15) & 100\\
 &  &  & Prop$_{(0.5)}$ & 75.88 (75.02, 76.71) & 74 & 62.27 (61.82, 62.84) & 36.8\\
 &  & 0.3 & HZ$_O$ & 77.20 (76.11, 78.26) & 100 & 65.74 (65.28, 66.17) & 100\\
 &  &  & Prop$_{(0.3)}$ & 74.95 (73.69, 76.01) & 69.4 & 62.17 (61.54, 64.68) & 42.3\\
\bottomrule
\end{tabular}
\begin{tablenotes}
\item 
  Patients denote the range of the number of patients.
  $S$ denotes the number of the population studies.
  $p$ denotes the approximate proportion of the published studies among the population.
	Median with 25th and 75th empirical quartiles (Q1, Q3) of the SAUC at $t=2$ are reported. 
	CR denotes the proportion of successfully convergenced estimates among 1000 repetition.
	HZ$_P$ denotes the HZ model using the population studies; 
	HZ$_O$ denotes the HZ model using only the corresponding numbers of published studies;
	Prop$_{(p)}$ denotes the proposed sensitivity analysis method given $p$.
	All the entries are multiplied by 100.
\end{tablenotes}
\end{threeparttable}
\end{table}

\clearpage
\appendix

\section{The asymptotic distribution of \texorpdfstring{$\hat{\boldsymbol{\lowercase{y}}}^{(\lowercase{i})}$}{yi} conditional on \texorpdfstring{${\boldsymbol \theta}^{(\lowercase{i})}$}{thetai} }\label{ap:sigma}

In this section, we are proving that $\hat{\boldsymbol y}^{(i)}=\left(\hat \mu_\se^{(i)}, \hat \mu_\sp^{(i)}, \hat \mu_\lnHR^{(i)}\right)'$ has the following asymptotic distribution, as mentioned in equation (\ref{eq:tnm1}):
\begin{align*}
\sqrt{n^{(i)}}
\left.
\begin{pmatrix}
\hat \mu_\se^{(i)}-\mu_\se^{(i)}\\ 
\hat \mu_\sp^{(i)}-\mu_\sp^{(i)}\\ 
\hat \mu_\lnHR^{(i)}-\mu_\lnHR^{(i)}
\end{pmatrix}
\right|
\begin{pmatrix}
\mu_\se^{(i)}\\ 
\mu_\sp^{(i)}\\ 
\mu_\lnHR^{(i)}
\end{pmatrix}
\overset{D}{\rightarrow}
N_3 \left( \mathbf 0, {\mathbf{\Sigma}}^{(i)} \right),
\text{~with~}
{\mathbf{\Sigma}}^{(i)} = 
\begin{bmatrix}
    \left\{\sigma_\se^{(i)}\right\}^2  &  \sigma_\mathrm{se,sp}^{(i)}  &  \sigma_\mathrm{se,lnHR}^{(i)} \\ 
       \sigma_\mathrm{se,sp}^{(i)}  & \left\{\sigma_\sp^{(i)}\right\}^2 &  \sigma_\mathrm{sp,lnHR}^{(i)} \\ 
     \sigma_\mathrm{se,lnHR}^{(i)}   &  \sigma_\mathrm{sp,lnHR}^{(i)}  & \left\{\sigma_\lnHR^{(i)}\right\}^2
\end{bmatrix}.
\end{align*}

By replacing the theoretical quantities in ${\mathbf{\Sigma}}^{(i)}$ with the corresponding consistent estimators, 
we can obtain the consistent estimator of ${\mathbf{\Sigma}}^{(i)}$, denoted by $\hat{\mathbf{\Sigma}}^{(i)}$.

\subsection{\texorpdfstring{$\mathbf H^{(i)}$}{Hi}: the asymptotic variance-covariance matrix of \texorpdfstring{$\left(\hat \mu_\se^{(i)}, \hat \mu_\sp^{(i)}\right)'$}{mu}}
\label{ap:A1}
Hattori and Zhou gave the proof of the asymptotic variance-covariance matrix of ${\left(\hat \mu_\se^{(i)}, \hat \mu_\sp^{(i)}\right)}'$ in their Appendices A and B. 
Following their notations, we define 
$x = S_1^{(i)}\left(t\right),~
y = S_0^{(i)}\left(t\right);~
z = q_1^{(i)} = P\left(\tilde X_j > v^{(i)}\right)$, and 
$w = q_0^{(i)}= P\left(\tilde X_j \le v^{(i)}\right)$.
Define
$
\mu^{(i)}_\se 
= \logit\left\{\se(v^{(i)}, t)\right\} 
= g_\se\left\{S_1^{(i)}\left(t\right), S_0^{(i)}\left(t\right), q_1^{(i)}, q_0^{(i)}\right\} 
= g^{(i)}_\se\left(x, y, z, w\right)
$ and, in the same way, 
$
\mu^{(i)}_\sp 
= g^{(i)}_\sp\left(x, y, z, w\right)
$.
It has been proved that
\begin{align*}
\sqrt{n^{(i)}}\left.
\begin{pmatrix}
\hat \mu_\se^{(i)}-\mu_\se^{(i)}\\ 
\hat \mu_\sp^{(i)}-\mu_\sp^{(i)}
\end{pmatrix}
\right|
\begin{pmatrix}
\mu_\se^{(i)}\\ 
\mu_\sp^{(i)}
\end{pmatrix}
\overset{D}{\rightarrow}
N_2 \left( \mathbf 0, \mathbf H^{(i)}\right),
\mathrm{~with~}
\mathbf H^{(i)}=
\begin{bmatrix}
    \left\{\sigma_\se^{(i)}\right\}^2       &  \sigma_\mathrm{se,sp}^{(i)}  \\ 
     \sigma_\mathrm{se,sp}^{(i)}   &  \left\{\sigma_\sp^{(i)}\right\}^2
\end{bmatrix},
\end{align*}
and
\begin{equation}
\begin{aligned}
\left\{\sigma_\se^{(i)}\right\}^2 
&= \left\{\dot{g}^{(i)}_{\se,x}\right\}^2\left(1/q_1^{(i)}\right) \left\{\sigma^{(i)}_1\left(t\right)\right\}^2
+ \left\{\dot{g}^{(i)}_{\se,y}\right\}^2\left(1/q_0^{(i)}\right) \left\{\sigma^{(i)}_0\left(t\right)\right\}^2 
+ \left\{\dot{g}^{(i)}_{\se,z} - \dot{g}^{(i)}_{\se,w}\right\}^2 q_1^{(i)}q_0^{(i)},\\
\left\{\sigma_\sp^{(i)}\right\}^2 
&= \left\{\dot{g}^{(i)}_{\sp,x}\right\}^2 \left(1/q_1^{(i)}\right) \left\{\sigma^{(i)}_1\left(t\right)\right\}^2
+ \left\{\dot{g}^{(i)}_{\sp,y}\right\}^2 \left(1/q_0^{(i)}\right) \left\{\sigma^{(i)}_0\left(t\right)\right\}^2 
+ \left\{\dot{g}^{(i)}_{\sp,z} - \dot{g}^{(i)}_{\sp,w}\right\}^2 q_1^{(i)}q_0^{(i)},\\
\sigma_\mathrm{se, sp}^{(i)}
& = \dot{g}^{(i)}_{\se,x}\dot{g}^{(i)}_{\sp,x} \left(1/q_1^{(i)}\right) \left\{\sigma^{(i)}_1\left(t\right)\right\}^2
+ \dot{g}^{(i)}_{\se,y}\dot{g}^{(i)}_{\sp,y} \left(1/q_0^{(i)}\right) \left\{\sigma^{(i)}_0\left(t\right)\right\}^2 
+ \left(\dot{g}^{(i)}_{\se,z} - \dot{g}^{(i)}_{\se,w}\right)
\left\{\dot{g}^{(i)}_{\sp,z} - \dot{g}^{(i)}_{\sp,w}\right\}q_1^{(i)}q_0^{(i)};
\end{aligned}
\label{eq:3sig}
\end{equation}
where
$\dot{g}^{(i)}_{\se,x}, \dot{g}^{(i)}_{\se,y}, \dot{g}^{(i)}_{\se,z}$, and $\dot{g}^{(i)}_{\se,w}$ 
are defined by the partial derivative of $g^{(i)}_\se(x, y, z, w)$ with respect to $x,y,z$, and $w$, respectively; 
$\dot{g}^{(i)}_{\sp,x}, \dot{g}^{(i)}_{\sp,y}, \dot{g}^{(i)}_{\sp,z}$, and $\dot{g}^{(i)}_{\sp,w}$ are defined in the same way for specificity;
$\sigma^{(i)}_1\left(t\right)$ and $\sigma^{(i)}_0\left(t\right)$ are the limiting variances of $\hat S_1^{(i)}\left(t\right)$ and $\hat S_0^{(i)}\left(t\right)$, respectively.
According to Section 3.2 in Hattori and Zhou,
$\sigma^{(i)}_1\left(t\right)$ and $\sigma^{(i)}_0\left(t\right)$ in equation (\ref{eq:3sig}) are consistently estimated by the Greenwood formula, in which the censoring distribution is estimated by using the median follow-up time. 
By replacing the theoretical quantities in equation (\ref{eq:3sig}) with their consistent estimators,
$\sigma_\se^{(i)}, \sigma_\sp^{(i)}$, and $\sigma_\mathrm{se, sp}^{(i)}$ can be consistently estimated, denoted by $\hat \sigma_\se^{(i)}, \hat \sigma_\sp^{(i)}$, and $\hat \sigma_\mathrm{se,sp}^{(i)}$, respectively, and the consistent estimator of $\mathbf H^{(i)}$ can be consequently obtained, denoted by $\hat{\mathbf H}^{(i)}$.

\subsection{\texorpdfstring{$\left\{\sigma_\lnHR^{(i)}\right\}^2$}{sigmailnHR2}: the asymptotic variance of \texorpdfstring{$\hat \mu_\lnHR^{(i)}$}{lnHR}}
\label{ap:A2}
In study $i$, we define $\tilde Z_j = \mathbf 1\left(\tilde X_j > v^{(i)}\right)$ for simplicity, where $\mathbf 1(\cdot)$ indicates the indicator function. 
Thus, $\tilde Z_j=1$ if subject $j$ belongs to the high expression group and $\tilde Z_j=0$ if the low expression group, 
and $\tilde Z_j\left(j=1, \dots,n^{(i)}\right)$ are the iid copies of $\tilde Z$.
Let $\hat \mu_\lnHR^{(i)}$ denote the maximum partial likelihood estimate from the Cox model 
\begin{align}
\lambda^{(i)}\left(t\right) = \lambda_0^{(i)}\left(t\right)\exp \left\{\mu_\lnHR^{(i)} \tilde Z \right\},
\label{eq:cox}
\end{align} 
where $\lambda^{(i)}\left(t\right)$ denotes the hazard function, $\lambda_0^{(i)}\left(t\right)$ the baseline hazard function.

Recall that $\tilde Y_j= \mathrm{min}\left(\tilde T_j, \tilde C_j\right)$ is the follow-up time; 
define $\tilde \Delta_j= \mathbf 1\left(\tilde T_j \le \tilde C_j\right)$, where $\left(\tilde Y_j, \Delta_j\right)$ are iid copies of $\left(\tilde Y, \tilde\Delta\right)$;
Define $N_j(u) = \mathbf 1\left(\tilde Y_j\le u, \tilde \Delta_j = 1\right)$ the counting process,
with $N_{0,j}(u) = \mathbf 1\left(\tilde Y_j\le u, \tilde \Delta_j = 1, \tilde X_j \le v^{(i)}\right)$ and $N_{1,j}(u) = \mathbf 1\left(\tilde Y_j\le u, \tilde \Delta_j = 1, \tilde X_j > v^{(i)}\right)$;
define $Y_j(u) = \mathbf 1\left(\tilde Y_j \ge u\right)$ the at-risk process,
with $Y_{0,j}(u) = \mathbf 1\left(\tilde Y_j \ge u, \tilde X_j \le v^{(i)}\right)$ and $Y_{1,j}(u) = \mathbf 1\left(\tilde Y_j \ge u, \tilde X_j > v^{(i)}\right)$,
and $0\le u\le t$.
The partial likelihood score function of the Cox model (\ref{eq:cox}) is 
\begin{align}
\mathbf U\left(\mu_\lnHR^{(i)}\right) &= \sum_{j = 1}^{n^{(i)}} \int_{0}^{t} \tilde Z_j 
-\dfrac{E\left\{\tilde Z Y(u)e^{\mu_\lnHR^{(i)} \tilde Z}\right\}}{E\left\{Y(u)e^{\mu_\lnHR^{(i)} \tilde Z}\right\}} dM_{\mathrm{cox},j}^{(i)}(u) \label{eq:score1}
\end{align}
with 
\begin{equation}
\begin{aligned}
dM_{\mathrm{cox},j}^{(i)}(u) &= dN_j(u)- Y_j(u)\lambda^{(i)}_j(u) 
= \tilde Z_jdM_{1,j}^{(i)}(u) + \left(1-\tilde Z_j\right)dM_{0,j}^{(i)}(u),\\
dM^{(i)}_{l,j}(u) &= dN_{l,j}(u) + Y_{l,j}(u)d\log S_l^{(i)}(u)
\end{aligned}
\label{eq:Mcox}
\end{equation}
where $l=0$ or $1$ indicates the low or high expression group;
$M_{\mathrm{cox},j}^{(i)}(u), M^{(i)}_{0,j}(u)$, and $M^{(i)}_{1,j}(u)$ indicates the martingales of the Cox model, the low, and the high expression group, respectively.

Let $\mathbf I\left(\mu_\lnHR^{(i)}\right)$ denote the information matrix,
and it holds that 
\begin{align}
\left\{n^{(i)}\right\}^{-1/2}\mathbf U\left(\mu_\lnHR^{(i)}\right) \overset{D}{\rightarrow} N\left(0, \mathbf I\left(\mu_\lnHR^{(i)}\right)\right).
\label{eq:score.pdf}
\end{align} 
Denote the sample information at $\mu_\lnHR^{(i)}$ to be $\hat{\mathbf I}\left(\mu_\lnHR^{(i)}\right)$, 
and it can be shown that $\left.\hat{\mathbf I}\left(\mu_\lnHR^{(i)}\right)\right/n^{(i)}\overset{P}{\rightarrow} \mathbf I\left(\mu_\lnHR^{(i)}\right)$.
To find the asymptotic normality of $\hat \mu_\lnHR^{(i)}$, 
we expand $\mathbf U\left(\hat \mu_\lnHR^{(i)}\right)$ about the true value $ \mu_\lnHR^{(i)}$:
\begin{align*}
\mathbf U\left(\hat \mu_\lnHR^{(i)}\right) \simeq \mathbf U\left(\mu_\lnHR^{(i)}\right) - 
\hat{\mathbf I}\left(\mu_\mathrm{lnHR*}^{(i)}\right) \left\{\hat \mu_\lnHR^{(i)} - \mu_\lnHR^{(i)}\right\}
\end{align*}
where $\left|\mu_{\lnHR}^{(i)} - \mu_\mathrm{lnHR*}^{(i)}\right|\le \left|\mu_\lnHR^{(i)} - \hat\mu_\lnHR^{(i)}\right|$.
Since $\mathbf U\left(\hat \mu_\lnHR^{(i)}\right) = 0$, we can obtain
\begin{align}
\sqrt{n^{(i)}}\left\{\hat \mu_\lnHR^{(i)} - \mu_\lnHR^{(i)}\right\} 
\simeq \left\{\left.\hat{\mathbf I}\left(\mu_\mathrm{lnHR*}^{(i)}\right)\right/n^{(i)}\right\}^{-1} \left\{n^{(i)}\right\}^{-1/2}\mathbf U\left(\mu_\lnHR^{(i)}\right).
\label{eq:Wi}
\end{align} 
Since $\hat \mu_\lnHR^{(i)} \overset{P}{\rightarrow} \mu_\lnHR^{(i)}$, 
which implies $\mu_\mathrm{lnHR*}^{(i)} \overset{P}{\rightarrow} \mu_\lnHR^{(i)}$,
and $\left.\hat{\mathbf I}\left(\mu_\mathrm{lnHR*}^{(i)}\right)\right/{n^{(i)}} \overset{P}{\rightarrow}  \mathbf I\left(\mu_\lnHR^{(i)}\right)$;
with (\ref{eq:score.pdf}) and Slutsky's theorem, it holds that
\begin{align}
\sqrt{n^{(i)}}\left\{\hat \mu_\lnHR^{(i)} - \mu_\lnHR^{(i)}\right\} \overset{D}{\rightarrow} N\left(0, \left\{\sigma_\lnHR^{(i)}\right\}^2 \right),
\mathrm{~with~}
\left\{\sigma_\lnHR^{(i)}\right\}^2 = \left\{\mathbf I\left(\mu_\lnHR^{(i)}\right)\right\}^{-1}.
\label{eq:pdf.lnhr}
\end{align}
Since $\left.\hat{\mathbf I}\left(\hat\mu_\lnHR^{(i)}\right)\right/n^{(i)}$ is also a consistent estimator of $\mathbf I\left(\mu_\lnHR^{(i)}\right)$, 
the asymptotic distribution in (\ref{eq:pdf.lnhr}) is unchanged if $\left\{\sigma_\lnHR^{(i)}\right\}^2$ is replaced by $\left\{\hat\sigma_\lnHR^{(i)}\right\}^2 = n^{(i)}\left\{\hat{\mathbf I}\left(\hat\mu_\lnHR^{(i)}\right)\right\}^{-1}$.
In practice, $\left\{\hat{\mathbf I}\left(\hat\mu_\lnHR^{(i)}\right)\right\}^{-1/2}$ can be observed as the SE of lnHR in the prognosis studies.

\subsection{\texorpdfstring{$\sigma_\mathrm{se,lnHR}^{(i)} $}{sigmaselnHR} and \texorpdfstring{$\sigma_\mathrm{sp,lnHR}^{(i)} $}{sigmasplnHR}: the asymptotic covariances between \texorpdfstring{$\left(\hat \mu_\se^{(i)}, \hat \mu_\lnHR^{(i)}\right)$}{Se,lnHR} and \texorpdfstring{$\left(\hat \mu_\sp^{(i)}, \hat \mu_\lnHR^{(i)}\right)$}{Sp,lnHR}}
\label{ap:A3}

In Appendix A of Hattori and Zhou, we knew that
\begin{align*}
\sqrt{n^{(i)}}\left\{\hat{\mu}^{(i)}_\se - \mu^{(i)}_\se\right\} 
\simeq \dot{g}^{(i)}_{\se,x}\sqrt{n^{(i)}/n_1^{(i)}}R_1^{(i)} + \dot{g}^{(i)}_{\se,y}\sqrt{n^{(i)}/n_0^{(i)}}R_0^{(i)} 
+\dot{g}^{(i)}_{\se,z}Q_1^{(i)} + \dot{g}^{(i)}_{\se,w}Q_0^{(i)}, 
\end{align*}
where $R_l^{(i)} = \sqrt{n_l^{(i)}}\left\{\hat{S}_l^{(i)}\left(t\right)-S_l^{(i)}\left(t\right)\right\}$
and $Q_l^{(i)} = \sqrt{n^{(i)}}  \left\{\hat{q}_l^{(i)} - q_l^{(i)}\right\}$ with $l=0,1$.

For simplicity, we denote $W^{(i)} = \sqrt{n^{(i)}}\left\{\hat \mu_\lnHR^{(i)} - \mu_\lnHR^{(i)}\right\}$,
and the integrand in (\ref{eq:score1}) is denoted by $\tilde Z_j -\boldsymbol{\mathcal{E}}$.
Thus, the covariance between $\left(\hat \mu_\se^{(i)}, \hat \mu_\lnHR^{(i)}\right)$ can be written
\begin{equation}
\begin{aligned}
&Cov\left(\sqrt{n^{(i)}}\left\{\hat{\mu}^{(i)}_\se - \mu^{(i)}_\se\right\},
\sqrt{n^{(i)}}\left\{\hat \mu^{(i)}_\lnHR - \mu^{(i)}_\lnHR\right\}\right) 
= Cov\left(\sqrt{n^{(i)}}\{\hat{\mu}^{(i)}_\se - \mu^{(i)}_\se\}, W^{(i)}\right)\\
\simeq &\dot{g}^{(i)}_{\se,x}\sqrt{n^{(i)}/n_1^{(i)}}Cov\left(R_1^{(i)}, W^{(i)}\right) 
+ \dot{g}^{(i)}_{\se,y}\sqrt{n^{(i)}/n_0^{(i)}}Cov\left(R_0^{(i)}, W^{(i)}\right)\\
&+ \dot{g}^{(i)}_{\se,z}Cov\left(Q_1^{(i)}, W^{(i)}\right) 
+ \dot{g}^{(i)}_{\se,w}Cov\left(Q_0^{(i)}, W^{(i)}\right).
\end{aligned}
\label{eq:cov1}
\end{equation}

Given $S_1^{(i)}\left(t\right)>0$, Hattori and Zhou defined
\begin{align*}
R_1^{(i)} = \sqrt{n_1^{(i)}}\left\{\hat{S}_1^{(i)}\left(t\right)-S_1^{(i)}\left(t\right)\right\}
\simeq \dfrac{1}{\sqrt{n_1^{(i)}}}\sum_{i=1}^{n_1^{(i)}} 
\left\{-S_1^{(i)}\left(t\right) \int_{0}^{t}\frac{\hat S_1^{(i)}(u-)}{S_1^{(i)}(u)}\frac{dM^{(i)}_{1,j}(u)}{E\left\{Y_{1}(u)\right\}}\right\}.
\end{align*}
According to (\ref{eq:score.pdf})-(\ref{eq:pdf.lnhr}) and define $B^{(i)} = \left.\hat{\mathbf I}\left(\hat\mu_\lnHR^{(i)}\right)\right/n^{(i)}$, we can write 
\begin{align*}
W^{(i)} = \sqrt{n^{(i)}}\left\{\hat \mu_\lnHR^{(i)} - \mu_\lnHR^{(i)}\right\}
\simeq \dfrac{1}{B^{(i)}} \dfrac{1}{\sqrt{n^{(i)}}} \sum_{j = 1}^{n^{(i)}} \int_{0}^{t} \tilde Z_j -\boldsymbol{\mathcal{E}} dM_{\mathrm{cox},j}^{(i)}(u), 
\end{align*} 
where, both $R_1^{(i)}$ and $W^{(i)}$ are local square integrable martingales. 
Thus, we can write
\begin{align*}
Cov\left(R_1^{(i)}, W^{(i)}\right) = E\left(R_1^{(i)} W^{(i)}\right )
\simeq  \dfrac{-S_1^{(i)}\left(t\right)}{B^{(i)}} \dfrac{1}{\sqrt{n_1^{(i)}n^{(i)}}}
\sum_{j=1}^{n^{(i)}}E \left\{\int_{0}^{t} 
\dfrac{\tilde Z -\boldsymbol{\mathcal E}}{E\left\{ Y_1^{(i)}(u)\right\}}d\left \langle   M^{(i)}_{1}, M_{cox}^{(i)}\right \rangle(u)\right\}.
\end{align*}
From (\ref{eq:Mcox}), we can obtain
\begin{align*}
 d\left \langle  M^{(i)}_{1,j}, M_{cox,j}^{(i)}\right \rangle(u)  
& \approx  d\left \langle M^{(i)}_{1,j}, \tilde Z_j dM_{1,j}^{(i)} + \left(1-\tilde Z_j\right)M_{0,j}^{(i)}\right \rangle(u) \\
 &=\tilde Z_j d\left \langle M^{(i)}_{1,j}, M^{(i)}_{1,j} \right \rangle(u) \\
 &= \tilde Z_j Y_j(u) d E\Lambda_1^{(i)}(u) \\
 &= Y_{1,j}(u) d\Lambda_1^{(i)}(u).~~~~~~~~~~\left(\because \tilde Z_jY_{j}(u) = Y_{1,j}(u)\right)
\end{align*}
Therefore, 
\begin{align*}
Cov\left(R_1^{(i)}, W^{(i)}\right)
&\simeq  \dfrac{-S_1^{(i)}\left(t\right)}{B^{(i)}} \dfrac{1}{\sqrt{n_1^{(i)}n^{(i)}}}
    \sum_{j=1}^{n^{(i)}}\int_{0}^{t} 
    E\left\{\dfrac{\tilde Z -\boldsymbol{\mathcal E}}{E\left\{ Y_{1}(u)\right\}} Y_{1}(u)\right\} d\Lambda_1^{(i)}(u)\\
&\simeq \dfrac{-S_1^{(i)}\left(t\right)}{B^{(i)}} \sqrt{\dfrac{n^{(i)}}{n_1^{(i)}}}\int_{0}^{t}
E \left \{ \dfrac{E\left\{\tilde ZY_{1}(u) -\boldsymbol{\mathcal E}Y_{1}(u)\right\}}{E\left\{ Y_1^{(i)}(u)\right\}} \right \} d\Lambda_1^{(i)}(u)\\
&= \dfrac{-S_1^{(i)}\left(t\right)}{B^{(i)}} \sqrt{\dfrac{n^{(i)}}{n_1^{(i)}}}\int_{0}^{t}1 - \boldsymbol{\mathcal E} d\Lambda_1^{(i)}(u)
~~~~~~~~~~~~~~~~~~~\left(\because \tilde Z_jY_{1,j}(u) = Y_{1,j}(u)\right)
\\
&\simeq\dfrac{S_1^{(i)}\left(t\right)}{B^{(i)}} \sqrt{\dfrac{n^{(i)}}{n_1^{(i)}}}\left\{
\log S_1^{(i)}\left(t\right) -\int_{0}^{t} \dfrac{1}{S^{(i)}(u)} d S_1^{(i)}(u)
\right\}
~~~~\left(\because \boldsymbol{\mathcal E} = \dfrac{S_1^{(i)}(u)}{S^{(i)}(u)}\right).
\end{align*}

Similarly, we can obtain 
\begin{align*}
Cov\left(R_0^{(i)}, W^{(i)}\right)
\simeq\dfrac{S_0^{(i)}\left(t\right)}{B^{(i)}} \sqrt{\dfrac{n^{(i)}}{n_0^{(i)}}}\left\{
\log S_0^{(i)}\left(t\right) -\int_{0}^{t} \dfrac{1}{S^{(i)}(u)} d S_0^{(i)}(u)
\right\}
\end{align*} 

Recall that $Q_1^{(i)} = \sqrt{n^{(i)}}  \left\{\hat{q}_1^{(i)} - q_1^{(i)}\right\}\simeq \sqrt{n^{(i)}} E\left\{\tilde Z - q_1^{(i)}\right\}$.
To find the covariance between $Q_1^{(i)}$ and $W^{(i)}$, we can write
\begin{align*}
Cov\left(Q_1^{(i)}, W^{(i)}\right)& =E\left(Q_1^{(i)} W^{(i)}\right) \\
& \simeq \dfrac{1}{n^{(i)}}\frac{1}{B^{(i)}}\sum_{j=1}^{n^{(i)}}
E \left\{ 
 E\left\{\tilde Z-q_1^{(i)} \right \}
\int_{0}^{t}\tilde Z - \boldsymbol{\mathcal E} dM_{cox,j}^{(i)}(u)\right\}\\
& \simeq \dfrac{1}{B^{(i)}}
E \left[
 E\left\{\tilde Z-q_1^{(i)} \right \}
E \left\{\int_{0}^{t}\tilde Z - \boldsymbol{\mathcal E} dM_{cox,j}^{(i)}(u)\right\}\right]\\
& = 0
\end{align*}
Similarly, $Cov\left(Q_0^{(i)}, W^{(i)}\right) \rightarrow 0 \text{ as } n^{(i)} \rightarrow \infty$.

Finally, we let $\sigma_\mathrm{se,lnHR}^{(i)}$ denote the asymptotic covariances (\ref{eq:cov1}) and it can be written by 
\begin{align}
\sigma_\mathrm{se,lnHR}^{(i)} 
&= {\dot{g}^{(i)}_{\se,x}}\dfrac{n^{(i)}S_1^{(i)}\left(t\right)}{n_1^{(i)}B^{(i)}} \left\{
\log S_1^{(i)}\left(t\right) -\int_{0}^{t} \dfrac{1}{S^{(i)}(u)} d S_1^{(i)}(u) \right\} 
+ {\dot{g}^{(i)}_{\se,y}}\dfrac{n^{(i)}S_0^{(i)}\left(t\right)}{n_0^{(i)}B^{(i)}} \left\{
\log S_0^{(i)}\left(t\right) -\int_{0}^{t} \dfrac{1}{S^{(i)}(u)} d S_0^{(i)}(u) \right\}.
\label{eq:sig13}
\end{align}
In the similar way, let $\sigma_\mathrm{sp,lnHR}^{(i)}$ denote the asymptotic covariances between $\left(\hat \mu_\sp^{(i)}, \hat \mu_\lnHR^{(i)}\right)'$:
\begin{align}
\sigma_\mathrm{sp,lnHR}^{(i)} 
= {\dot{g}^{(i)}_{\sp,x}}\dfrac{n^{(i)}S_1^{(i)}\left(t\right)}{n_1^{(i)}B^{(i)}} \left\{
\log S_1^{(i)}\left(t\right) -\int_{0}^{t} \dfrac{1}{S^{(i)}(u)} d S_1^{(i)}(u) \right\} 
+ {\dot{g}^{(i)}_{\sp,y}}\dfrac{n^{(i)}S_0^{(i)}\left(t\right)}{n_0^{(i)}B^{(i)}} \left\{
\log S_0^{(i)}\left(t\right) -\int_{0}^{t} \dfrac{1}{S^{(i)}(u)} d S_0^{(i)}(u) \right\}.
\label{eq:sig23}
\end{align}
By replacing $S_l^{(i)}$ with the observed KM estimators $\hat S_l^{(i)}~(l=0,1)$,
we can obtain the consistent estimators of $\sigma_\mathrm{se,lnHR}^{(i)}$ and $\sigma_\mathrm{sp,lnHR}^{(i)}$, 
denoted by $\hat \sigma_\mathrm{se,lnHR}^{(i)}$ and $\hat \sigma_\mathrm{sp,lnHR}^{(i)}$, respectively.
The integrations in the asymptotic covariances can be approximated by the trapezoidal rule: 
\begin{align*}
\int_{0}^{t} \frac{1}{S^{(i)}(u)} d S_l^{(i)}(u)
\approx \sum_{k =1}^{K}\left[ 
\frac{1}{\hat S^{(i)}\left(t_k\right)} + \frac{1}{\hat S^{(i)}\left(t_{k-1}\right)}
\right]
\frac{\hat S_l^{(i)}\left(t_k\right) - \hat S_l^{(i)}\left(t_{k-1}\right)}{2}
\end{align*}
where
\begin{align*}
{\hat S^{(i)}\left(t\right)} &= \frac{n_0^{(i)}\hat S_0^{(i)}\left(t\right) + n_1^{(i)}\hat S_1^{(i)}\left(t\right)}{n^{(i)}}.
\end{align*}

\section{The confidence interval of the time-dependent SAUC}
\label{ap:sauc}

The variance of the time-dependent SAUC, denoted SAUC$(t)$, can be constructed by the delta method.
Define $\boldsymbol D = \int_0^1\SROC(t)\left\{1-\SROC(t)\right\}\nabla \SROC(t)$,
where $\nabla \SROC(t)$ is the gradient of the linear function in SROC$(t)$ with 
$$\nabla \SROC(t) = \left(1, -\rho_1\tau_\se/ \tau_\sp,
-\rho_1/\tau_\sp\left\{\logit(x)+\mu_\se\right\}, 
-\rho_1\tau_\se/\tau^2_\sp\left\{\logit(x)+\mu_\se\right\},  
-\tau_\se/\tau_\sp\left\{\logit(x)+\mu_\se\right\}
\right)'.$$
Let $\hat{\boldsymbol D}$ denote the MLE of $\boldsymbol D$ by replacing the unknown parameters with their MLEs.
By the delta method, the variance of the estimated SAUC$(t)$ is consistently estimated by 
\begin{align*}
Var\left[\mathrm{SA\hat UC}(t)\right] = \hat{\boldsymbol D}' \hat{\boldsymbol \Upsilon}\hat{\boldsymbol D}.
\end{align*}
For the HZ model (\ref{eq:bnm12}), $\hat{\boldsymbol \Upsilon}$ is the estimated variance-covariance matrix of $\hat{\boldsymbol \mu}$ and $\hat{\boldsymbol \Omega}$. For the proposed method, $\hat{\boldsymbol \Upsilon}$ is the estimated variance-covariance matrix of $\hat{\boldsymbol \theta}$ and $\hat{\boldsymbol \Psi}$.
By applying the delta method to the logit-transformed SAUC$(t)$, the two-tailed confidence interval of the $\mathrm{SA\hat UC}(t)$ at significance level $\eta$ is estimated by
\begin{align*}
\logit^{-1}\left\{\logit\left(\mathrm{SA\hat UC}(t)\right)\pm z_{1-\eta/2}\dfrac{ \sqrt{Var\left[\mathrm{SA\hat UC}(t)\right]}}{\mathrm{SA\hat UC}(t)\left[1-\mathrm{SA\hat UC}(t)\right]}\right\}.
\end{align*}

\clearpage

\section*{Supplementary Material for ``Sensitivity analysis for publication bias on the time-dependent summary ROC analysis in meta-analysis of prognostic studies''}

\section*{1 The estimations of the SROC\texorpdfstring{$(\lowercase{t})$}{(t)} and the SAUC\texorpdfstring{$(\lowercase{t})$}{(t)} in Ki67 example}

As mentioned in Section 5, we present the other estimated parameters of the SROC$(t)$ at $t=3$ by the HZ model ($p=1$) and the proposed method $(p=0.6, 0.4, 0.2)$ in Table \ref{tab:sroc-ki67-3} and those at $t=5$ in Table \ref{tab:sroc-ki67-5}.
The estimates of the SAUC$(t)$ at $t=3$ or $t=5$ by the the HZ model ($p=1$) and the proposed method $(p=0.9,\dots, 0.1)$ are presented in Table \ref{tab:sauc-ki67}.

\section*{2 The scenarios used in the simulation}
As mentioned in Section 6, we presented the values of $\alpha$ in Table \ref{tab:set}.




\begin{table}[!htb]

\caption{\label{tab:sroc-ki67-3}Estimates of the other parameters for the SROC$(3)$}
\centering
\begin{threeparttable}
\begin{tabular}[t]{rrrrrrrrrrrr}
\toprule
$p$ & $\mu_\mathrm{se}$ (se) & $\mu_\mathrm{sp}$ (sp) & $\mu_\mathrm{lnHR}$ (HR) & $\psi_\mathrm{se}$ & $\psi_\mathrm{sp}$ & $\psi_\mathrm{lnHR}$ & $\rho_1$ & $\rho_2$ & $\rho_3$ & $\beta$ & $\alpha$\\
\midrule
1* & 0.670 (0.662) & 0.282 (0.570) &  & 0.705 & 0.51 &  & -0.855 &  &  &  & \\
0.6 & 0.688 (0.666) & 0.204 (0.551) & 0.452 (1.571) & 0.724 & 0.514 & 0.493 & -0.851 & 0.442 & -0.061 & 1.692 & -1.23\\
0.4 & 0.726 (0.674) & 0.123 (0.531) & 0.231 (1.260) & 0.72 & 0.522 & 0.545 & -0.846 & 0.497 & -0.129 & 1.569 & -1.684\\
0.2 & 0.828 (0.696) & -0.016 (0.496) & -0.110 (0.896) & 0.719 & 0.534 & 0.598 & -0.848 & 0.559 & -0.229 & 1.506 & -2.1\\
\bottomrule
\end{tabular}
\begin{tablenotes}
\item 
			$\rho_1$ denotes the correlation coefficient between $\mu_\mathrm{sen}$ and $\mu_\mathrm{spe}$;
			$\rho_2$ denotes that between $\mu_\mathrm{sen}$ and $\mu_\mathrm{lnHR}$;
			$\rho_3$ denotes that between $\mu_\mathrm{spe}$ and $\mu_\mathrm{lnHR}$.
			$^*p=1$ indicates the HZ model without considering PB.
\end{tablenotes}
\end{threeparttable}
\end{table}
 
\begin{table}[!htb]

\caption{\label{tab:sroc-ki67-5}Estimates of the other parameters for the SROC$(5)$}
\centering
\begin{threeparttable}
\begin{tabular}[t]{rrrrrrrrrrrr}
\toprule
$p$ & $\mu_\mathrm{se}$ (se) & $\mu_\mathrm{sp}$ (sp) & $\mu_\mathrm{lnHR}$ (HR) & $\psi_\mathrm{se}$ & $\psi_\mathrm{sp}$ & $\psi_\mathrm{lnHR}$ & $\rho_1$ & $\rho_2$ & $\rho_3$ & $\beta$ & $\alpha$\\
\midrule
1* & 0.526 (0.629) & 0.354 (0.588) &  & 0.444 & 0.58 &  & -0.938 &  &  &  & \\
0.6 & 0.510 (0.625) & 0.264 (0.566) & 0.417 (1.517) & 0.461 & 0.585 & 0.482 & -0.897 & 0.406 & 0.107 & 1.734 & -1.18\\
0.4 & 0.520 (0.627) & 0.174 (0.543) & 0.194 (1.214) & 0.458 & 0.594 & 0.54 & -0.895 & 0.468 & 0.014 & 1.588 & -1.603\\
0.2 & 0.574 (0.640) & 0.014 (0.503) & -0.149 (0.862) & 0.456 & 0.607 & 0.598 & -0.903 & 0.541 & -0.116 & 1.51 & -1.996\\
\bottomrule
\end{tabular}
\begin{tablenotes}
\item 
			$\rho_1$ denotes the correlation coefficient between $\mu_\mathrm{sen}$ and $\mu_\mathrm{spe}$;
			$\rho_2$ denotes that between $\mu_\mathrm{sen}$ and $\mu_\mathrm{lnHR}$;
			$\rho_3$ denotes that between $\mu_\mathrm{spe}$ and $\mu_\mathrm{lnHR}$.
			$^*p=1$ indicates the HZ model without considering PB.
\end{tablenotes}
\end{threeparttable}
\end{table}
 
\begin{table}[!htb]

\caption{\label{tab:sauc-ki67}The estimated SAUC$(t)$ with 95\% condifence intervals at $t=3,5$}
\centering
\begin{tabular}[t]{rrr}
\toprule
 & SAUC(3) (95\% CI) & SAUC(5) (95\% CI)\\
\midrule
HZ & 0.649 (0.606, 0.690) & 0.646 (0.610, 0.680)\\
$p=0.9$ & 0.647 (0.604, 0.687) & 0.644 (0.609, 0.678)\\
$p=0.8$ & 0.644 (0.601, 0.685) & 0.641 (0.605, 0.675)\\
$p=0.7$ & 0.641 (0.597, 0.683) & 0.636 (0.599, 0.672)\\
$p=0.6$ & 0.638 (0.591, 0.682) & 0.632 (0.592, 0.670)\\
$p=0.5$ & 0.634 (0.583, 0.682) & 0.627 (0.583, 0.669)\\
$p=0.4$ & 0.631 (0.574, 0.684) & 0.623 (0.573, 0.670)\\
$p=0.3$ & 0.627 (0.562, 0.688) & 0.618 (0.560, 0.672)\\
$p=0.2$ & 0.624 (0.546, 0.695) & 0.613 (0.545, 0.677)\\
$p=0.1$ & 0.621 (0.526, 0.708) & 0.608 (0.525, 0.685)\\
\bottomrule
\end{tabular}
\end{table}

\begin{table}[!htb]

\caption{\label{tab:set}Scenarios of simulation}
\centering
\begin{tabular}[t]{rrrrrrr}
\toprule
Censoring & Biomarker & Subjects & $\beta$ & $\alpha_{0.7}$ & $\alpha_{0.5}$ & $\alpha_{0.3}$\\
\midrule
$Exp(0.2)$ & 1 & $U(50,150)$ & 1 & -2.431 & -3.199 & -3.961\\
 &  & $U(50,300)$ & 1 & -3.277 & -4.178 & -5.083\\
 & 2 & $U(50,150)$ & 1 & -0.518 & -1.270 & -2.020\\
 &  & $U(50,300)$ & 1 & -0.878 & -1.649 & -2.427\\
\addlinespace
$U(1,4)$ & 1 & $U(50,150)$ & 1 & -2.968 & -3.713 & -4.452\\
 &  & $U(50,300)$ & 1 & -3.945 & -4.899 & -5.818\\
 & 2 & $U(50,150)$ & 1 & -0.795 & -1.547 & -2.286\\
 &  & $U(50,300)$ & 1 & -1.233 & -2.020 & -2.802\\
\bottomrule
\end{tabular}
\end{table}

\end{document}